\documentclass[
superscriptaddress,
final,
reprint,
showkeys,
pre,
aps,
floatfix,
]{revtex4-2}

\usepackage[dvips]{graphicx}
\usepackage[utf8]{inputenc}
\usepackage{amsmath,amssymb,booktabs,latexsym,MnSymbol,bm}
\newcommand\de{\mathrel{\bullet\mkern-2.5mu{\rightarrow}}}
\newcommand\ie{\mathrel{{\leftrightarrow}}}

\newcommand*\IPpp{\mbox{$I\!P\!\!+\!\!+$}}
\newcommand*\Ipp{\mbox{$I\!\!+\!\!+$}}
\newcommand*\Ppp{\mbox{$P\!\!+\!\!+$}}
\newcommand*\IpPp{\mbox{$I\!\!+\!\!P+$}}
\newcommand*\IpPm{\mbox{$I\!\!+\!\!P-$}}
\newcommand*\ImPp{\mbox{$I\!\!-\!\!P+$}}
\newcommand*\ImPm{\mbox{$I\!\!-\!\!P-$}}

\usepackage[%
  colorlinks=true,
  urlcolor=blue,
  linkcolor=blue,
  citecolor=blue
]{hyperref}
\usepackage{lipsum}

\usepackage[final]{pdfpages}
\makeatletter
\AtBeginDocument{\let\LS@rot\@undefined}
\makeatother

\begin{document}

\title{Association between productivity and journal impact across disciplines and career age}

\author{Andre\ S.\ Sunahara}
\affiliation{Departamento de F\'isica, Universidade Estadual de Maring\'a -- Maring\'a, PR 87020-900, Brazil}

\author{Matja{\v z} Perc} 
\email{matjaz.perc@gmail.com}
\affiliation{Faculty of Natural Sciences and Mathematics, University of Maribor, Koro{\v s}ka cesta 160, 2000 Maribor, Slovenia}
\affiliation{Department of Medical Research, China Medical University Hospital, China Medical University, Taichung 404332, Taiwan}
\affiliation{Alma Mater Europaea ECM, Slovenska ulica 17, 2000 Maribor, Slovenia}
\affiliation{Complexity Science Hub Vienna, Josefst{\"a}dterstra{\ss}e 39, 1080 Vienna, Austria}

\author{Haroldo V.\ Ribeiro}
\email{hvr@dfi.uem.br}
\affiliation{Departamento de F\'isica, Universidade Estadual de Maring\'a -- Maring\'a, PR 87020-900, Brazil}
\date{\today}

\begin{abstract}
The association between productivity and impact of scientific production is a long-standing debate in science that remains controversial and poorly understood. Here we present a large-scale analysis of the association between yearly publication numbers and average journal-impact metrics for the Brazilian scientific elite. We find this association to be discipline-specific, career-age dependent, and similar among researchers with outlier and non-outlier performance. Outlier researchers either outperform in productivity or journal prestige, but they rarely do so in both categories. Non-outliers also follow this trend and display negative correlations between productivity and journal prestige but with discipline-dependent intensity. Our research indicates that academics are averse to simultaneous changes in their productivity and journal-prestige levels over consecutive career years. We also find that career patterns concerning productivity and journal prestige are discipline-specific, having in common a raise of productivity with career age for most disciplines and a higher chance of outperforming in journal impact during early career stages.
\end{abstract}

\maketitle

\section*{Introduction}

The development of knowledge-based economies and the increasing availability of information and knowledge itself have driven transdisciplinary efforts towards a better quantitative understanding of the scientific enterprise: the science of Science~\cite{zeng2017science,fortunato2018science}. Beyond the academic question of finding driving mechanisms of Science, these initiatives aim to enhance scientific efficiency by identifying successful practices and policies, from the choice of countries' scientific priorities to the selection of research projects and faculty candidates. Scientific progress is nowadays strongly dependent on research evaluation processes, as they regulate the stream of ideas and research projects by means of science funding allocation~\cite{azoulay2009incentives,hua2016interdisciplinary,fortunato2018science,meirmans2019science}. But while peer review is considered the standard approach for assessing academic performance~\cite{wilsdon2016metric}, the process itself is laborious and has several drawbacks, ranging from biases and lack of consistency to fraud~\cite{wessely1998peer,smith2006peer,wilsdon2016metric,balietti2016peer}. In addition, the increasing number of scientific publications~\cite{bornmann2015growth} and the growth of the scientific workforce~\cite{ioannidis2014estimates} impose further limitations on the peer-review method~\cite{meirmans2019science}. A direct consequence of these issues is the steady increase (especially after the 2000s~\cite{cameron2005trends}) in the use of bibliometric indexes for grading the performance of researchers~\cite{traag2019systematic,meirmans2019science}.

Bibliometric assessments are considered more objective criteria, but there is no consensus on which indexes are more suitable for evaluating academic performance, and many believe that the intrinsic nature of scientific processes can only be precisely quantified by multidimensional features~\cite{gagolewski2013scientific,siudem2020three}. This data-driven culture of performance evaluation has amassed much criticism~\cite{dora,hicks2015bibliometrics,wilsdon2016metric,nuffield}, and it also poses enormous pressure on scholars (particularly on young scientists~\cite{powell2016junior}) for publishing in large quantities, in prestigious journals, and developing highly cited research~\cite{meirmans2019science,moher2018assessing,schimanski2018evaluation}. Still, research productivity and impact measures are often and widely used to quantify academic performance, representing essential ingredients for the perception and recognition of academic success. While productivity is defined as the number of research items in a given period, impact has a more subjective character and is usually measured by the number of citations, the share of articles among highly cited papers, and the prestige of the publication venue. Regardless of which metric is used, research evaluation via bibliometrics has raised the ``quality versus quantity'' debate since its conception~\cite{dennis1954productivity,white1978relation,lawani1986some,simonton1988scientific,feist1997quantity,haslam2010quality,nijstad2010dual,bosquet2013academics,abramo2014authors,sandstrom2016quantity,lariviere2016many,garousi2017quantity,michalska2017and,kolesnikov2018researchers,bornmann2019productivity,forthmann2020investigating}, and there is still no agreement on the association between productivity and impact. For instance, while Larivi{\`e}re and Costas~\cite{lariviere2016many} have found a positive association between productivity and number of highly cited articles, Bornmann and Tekles~\cite{bornmann2019productivity} have shown that top-productive authors usually have lower fractions of publications among top-cited articles (that is, a negative association between productivity and impact at overly high productivity levels). An important part of these controversial findings reflects the fact that the association between productivity and impact is discipline-specific, career-age dependent, scale-dependent, and may be affected by the presence of outlier individuals. However, there is still a lack of works simultaneously addressing all these points to reveal the overall complexity of the ``quantity versus quality'' relationship.
 
Here we investigate multifaceted aspects of this association by analyzing the scientific career of more than six thousand scientists from the Brazilian scientific community's elite from 14 different disciplines. We determine the yearly publication numbers and the respective average value of the journal-impact metrics over the careers of these academics. Although the use of journal-level metrics for assessing the individual performance of researchers is controversial~\cite{lariviere2019jifhistory,dora}, this approach remains widespread and largely used~\cite{mckiernan2019meta}, especially in Brazil where several universities use journal prestige (or derivative indicators) for everything from grading resumes of graduate students to the selection of tenure-track faculty positions and grant applications. Recent works have also demonstrated that journal-level metrics carry information about academic performance~\cite{bornmann2017skewness,traag2021inferring,kim2019scientific,correa2020scihub,waltman2020use} and that these metrics are correlated with citations, thus indicating that citations and journal-level metrics are partly substitutes. Whether journal-level metrics (or even citations) are suitable or not for research evaluation -- fact is that these metrics are still important for the scientific community and deserve further investigation. 

Our research probes patterns of the association between productivity and journal-impact metrics throughout researchers' careers across different disciplines. In contrast to previous works, we use standard score measures to account for discipline and inflation-like effects and correct for size-dependent biases on average journal prestige. We further identify outlier individuals in productivity and journal impact, finding these academics to either outperform in productivity or journal prestige over their careers but rarely in both categories. We also find that academics are averse to simultaneously changing their levels of productivity and journal prestige and prefer maintaining these levels over consecutive years of their careers. For non-outlier individuals, our results indicate a negative correlation between productivity and journal prestige for most researchers from most disciplines. However, we show that career patterns of productivity and journal prestige are discipline-specific, although they have in common the fact that productivity increases with time for all disciplines. By shedding light on career-age and discipline-specific aspects of productivity and journal prestige, we believe our work may contribute significantly to a more comprehensive and fair research evaluation process.

\section*{Results}

\subsection*{Journal prestige versus productivity plane}

To investigate the association between productivity and journal prestige, we have collected the academic curricula of 6,028 Brazilian researchers from 14 disciplines (see Methods for details) holding the CNPq Research Productivity Fellowship (\textit{Bolsa Produtividade em Pesquisa do CNPq}) as of May 2017. This fellowship has been awarded since the 1970s by the Brazilian National Council for Scientific and Technological Development (CNPq -- \textit{Conselho Nacional de Desenvolvimento Científico e Tecnológico}) in recognition to outstanding scientific production. CNPq fellows have significant status among the Brazilian scientific community and are often considered the elite of Brazilian scientists. We further obtain the Journal Impact Factor (JIF) between 1997 and 2015 from Clarivate's Journal Citation Reports. We combine these data sets to assign the time-varying values of JIF to the 312,881 articles published by the CNPq fellows between 1997 and 2015. We consider the number of articles published per year as the productivity indicator and the average JIF as a proxy for journal prestige. We have also carried out a comparative analysis when considering the Scopus' SCImago Journal Rank (SJR) as an indicator of journal prestige. Despite the substantial differences in the definitions of JIF and SJR, both measures of journal prestige are strongly correlated (Fig.~S1~\cite{SI}), and yield very similar results. We have opted to present the results for the JIF in the main text, and we refer to the Supplementary Material~\cite{SI} for comparisons with the SJR. 

We start our investigation by noticing that the number of articles and citations have increased over time~\cite{solla1963little,sinatra2016quantifying}. This produces inflation in productivity and journal-impact measures that needs to be accounted for a fair comparison between different publication years. Our results indicate that the average productivity of the CNPq fellows has increased at a rate $\approx1.57$ papers/year per decade. Similarly, the average JIF of these publications has raised $\approx0.72$ units per decade (Figs.~S2 and S3~\cite{SI}). This inflation effect is different among disciplines; for instance, the productivity of Medicine researchers has increased $\approx3.5$ papers/year per decade, while those working in Electrical Engineering experienced a productivity inflation of $\approx0.3$ papers/year per decade. Because of this inflation effect and differences in publication patterns among disciplines, we do not use raw numbers of productivity but instead robust standard scores ($z$-scores) relative to discipline and year of publication. In addition to discipline and year, the robust standard scores for average journal prestige are also relative to researchers' productivity levels. This additional normalization accounts for the fact that the more productive a researcher is in a given year, the narrower the range of variation of his/her average journal prestige. A similar size effect has been observed by Antonoyiannakis~\cite{antonoyiannakis2018impact,antonoyiannakis2020impact} when comparing the impact factor of journals with different sizes, and the approach we use for rescaling the average journal prestige is an adapted version of his method for ranking journals~\cite{antonoyiannakis2018impact,antonoyiannakis2020impact}.

\begin{figure*}[!ht]
  \centering
  \includegraphics[width=1\textwidth, keepaspectratio]{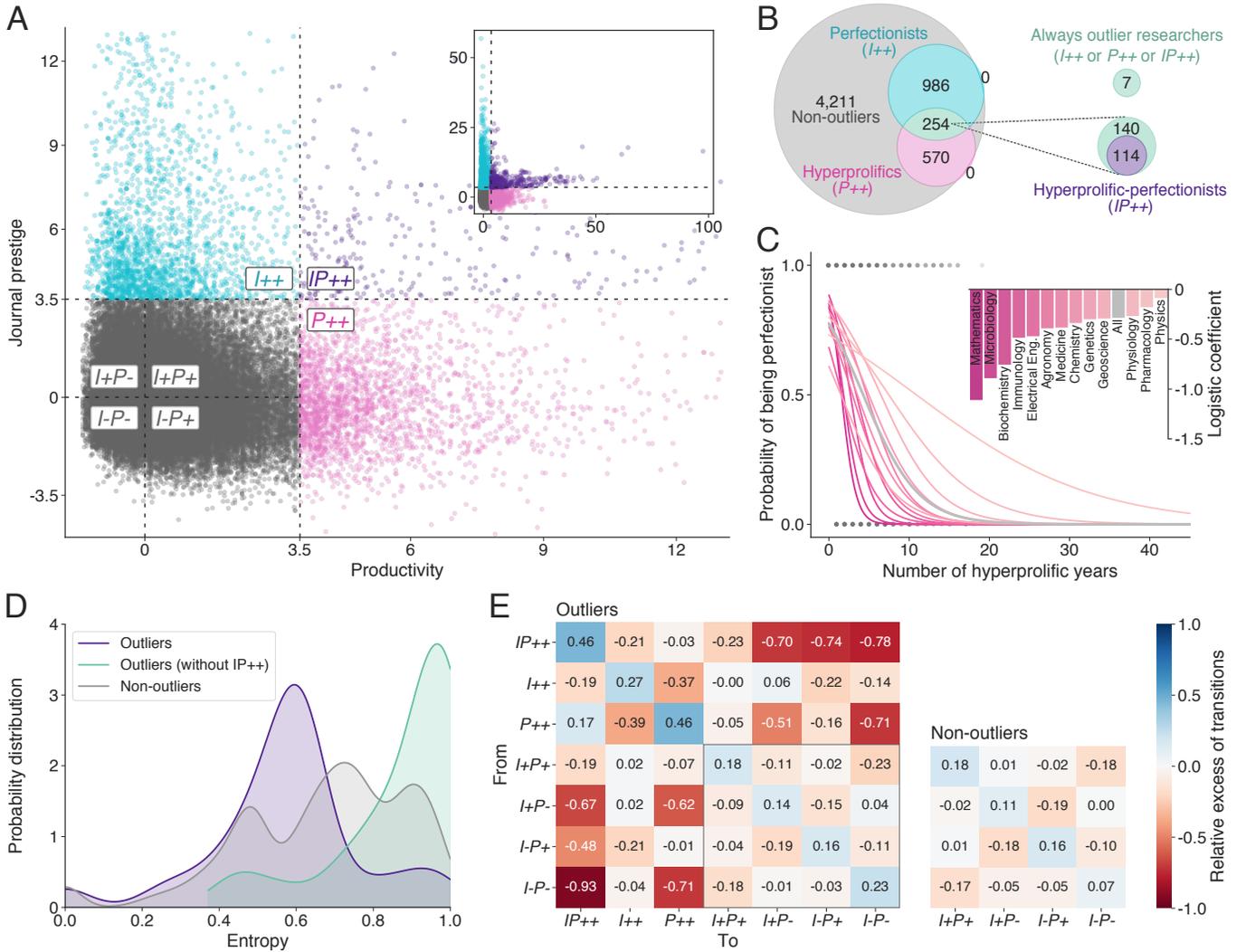}
  \caption{\textbf{Journal prestige versus productivity.} (A) Relation between average journal impact and productivity in standard score units (the inset shows the full range of the plane). Data points represent career years of researchers from the 14 disciplines in our study. This plane is divided into seven sectors. Three sectors represent career years with overly high performance in journal impact (\Ipp{}), productivity (\Ppp{}), or both quantities (\IPpp{}). Four non-outlier sectors represent career years with productivity and journal impact above (\IpPp{}) or below (\ImPm{}) the average, journal impact below and productivity above the average (\ImPp{}), and journal impact above and productivity below the average (\IpPm{}). (B) Venn diagram showing the set relations among the four categories of researchers. Non-outliers are those with all career years in non-outlier sectors. Perfectionists and hyperprolifics are researchers with at least one career year in sectors \Ipp{} and \Ppp{}, respectively. Hyperprolific-perfectionists are those having at least one career year within sector \IPpp{}. (C) Probability of being a perfectionist researcher while having a given number of career years in the hyperprolific sector (\Ppp{}), as estimated via logistic regression (the inset shows the logistic coefficients). The colored curves (and bars) refer to different disciplines, while the gray-colored curve represents the aggregated result of all disciplines. Materials Engineering (omitted in this panel) is the only discipline that does not display a significant association. (D) Probability distribution of the normalized entropy values associated with the occupation of the plane sectors over researchers' careers. The purple curve shows the results for the occupation of only outlier sectors by outlier researchers and the green curve is the same but after ignoring sector \IPpp{}. The gray curve shows the entropy distribution for non-outlier researchers. (E) Transition matrix among the plane sectors for outlier (left) and non-outlier (right) researchers. Each cell represents the relative excess of transitions between two sectors compared with a null model corresponding to shuffled versions of researchers' careers for 10,000 realizations.}
  \label{fig:1}
\end{figure*}

Figure~\ref{fig:1}A shows a scatter plot of the average journal prestige versus the productivity for all career years of researchers in our data set (see Fig.~S4~\cite{SI} for a comparison with the SJR data). In this plane, a unit of productivity indicates a performance one standard deviation above (if positive) or below (if negative) the average performance of all scholars of a given discipline in a given year. Similarly, a unit of average journal prestige represents a performance one standard deviation above (if positive) or below (if negative) the average random performance at a given productivity level of a given discipline and year. We divide this plane into four main sectors separating outlier years of researchers ($z$-scores higher than $3.5$) regarding productivity ($P$) and average journal impact ($I$). The sector \IPpp{} contains career years in which researchers were simultaneously outliers in journal prestige and productivity ($I>3.5$ and $P>3.5$). Similarly, sectors \Ipp{} and \Ppp{} indicate outlier career years only regarding journal prestige ($I>3.5$ and $P<3.5$) and productivity ($I<3.5$ and $P>3.5$), respectively. We further divide the non-outlier sector ($I<3.5$ and $P<3.5$) into four other sectors: 
\IpPp{} for career years with journal prestige and productivity above the average ($I>0$ and $P>0$); 
\IpPm{} for career years with journal prestige above and productivity below the average ($I>0$ and $P<0$); 
\ImPp{} for career years with journal prestige below and productivity above the average ($I<0$ and $P>0$); 
and \ImPm{} for career years with journal prestige and productivity below the average ($I<0$ and $P<0$). 

\subsection*{Outlier and non-outlier researchers}

One of the most striking features of the plane shown in Fig.~\ref{fig:1}A is the existence of researchers who, despite belonging to the elite of Brazilian scientists, further differentiate themselves by exhibiting productivity or average journal prestige (or both) in overly high levels at specific years of their careers. These outlier career years are relatively rare and represent only 7.7\% of the 76,454 total career years (Fig.~S5A~\cite{SI}). Among the outlier sectors, the number of career years in \Ppp{} and \Ipp{} represent 47\% and 46\% of the total, respectively. Consequently, career years in sector \IPpp{} are much rarer and correspond to only 7\% of the total outlier years. Similar results are obtained for the SJR data set (Fig.~S5B~\cite{SI}).

Outlier years also represent only a small fraction of the careers of researchers covered by our data set (Fig.~S6A~\cite{SI}). More than 47.6\% of these researchers are outliers in productivity or journal prestige (or both) only in one year, and only 6.7\% have more than 50\% of their career years in outlier sectors. The Venn diagram of Fig.~\ref{fig:1}B depicts the set relations between researchers categorized as non-outlier (all career years in non-outlier sectors), perfectionist (at least one career year in sector \Ipp{}), hyperprolific (at least one career year in sector \Ppp{}), and hyperprolific-perfectionist (at least one career year in sector \IPpp{}). About 30\% of all researchers manage to have at least one career year in outlier sectors. There is no researcher with all career years in sector \IPpp{} nor either in sector \Ipp{} or \Ppp{}. In addition, only seven researchers (a chemist, an agronomist, and five physicists) have all career years covered by our data set in the three outlier sectors. Similar results are found for the SJR data set (Fig.~S6B~\cite{SI}).

Among the 1,817 outlier researchers, 1,556 (85.6\%) are only hyperprolifics or only perfectionists over their careers. This result indicates that most outlier researchers have a persistent behavior regarding being hyperprolific or perfectionist. This clear distinction between hyperprolifics and perfectionists is further corroborated by the existence of only 121 researchers (6.7\% of the outliers) simultaneously outliers in both categories, that is, in sector \IPpp{}. A similar pattern was recently observed by Bornmann and Tekles~\cite{bornmann2019productivity} for the association between productivity and number of articles in the top-1\% most cited. Our result thus indicates that it is extremely hard to frequently publish in very prestigious journals and keep productivity at overly high levels. Intriguingly, we observe that extremely hyperprolific research years ($P>27.7$) are all in sector \IPpp{}. This result shows that while very rare, there are sixteen researchers capable of maintaining extreme performances in productivity and journal prestige.

To reinforce this result, we use logistic regression to estimate the effect of hyperprolific years on the probability of being a perfectionist researcher (see Methods for details). Figure~\ref{fig:1}C shows the probability of being a perfectionist researcher as a function of the number of hyperprolific years and the logistic coefficients when considering all disciplines both together and separated. Materials Engineering does not show a significant association ($p\text{-value}>0.05$) and has been omitted in Fig.~\ref{fig:1}C. For the other thirteen disciplines and when aggregating all disciplines, the coefficients are significant and negative, establishing that an increase in the number of hyperprolific years decreases the chances of being a perfectionist researcher. However, this effect varies considerably among the disciplines. For instance, while five hyperprolific years practically prevent the existence of perfectionist researchers in Mathematics, there is a probability of 63.2\% of being a perfectionist in Physics with the same number of hyperprolific years. For the SJR data set, 23 out of 25 disciplines display a negative and significant association between the number of hyperprolific years and the probability of being a perfectionist (Fig.~S4C~\cite{SI}), reaffirming the negative association between these two behaviors.

The group of 261 researchers who manage to publish both as perfectionist and hyperprolific (simultaneously or not) is significantly more productive than the exclusively hyperprolific ones (average $z$-score productivity of $2.71\pm0.08$ versus $2.06\pm0.03$; $p$-value~$<10^{-16}$, permutation test) and the exclusively perfectionist ones (average $z$-score productivity of $2.71\pm0.08$ versus $0.54\pm0.02$; $p$-value~$<10^{-16}$, permutation test). Furthermore, this former group of researchers publish in journals with higher prestige than hyperprolific (average $z$-score JIF of $1.89\pm0.05$ versus $0.23\pm0.02$; $p$-value~$<10^{-16}$, permutation test) and perfectionist researchers (average $z$-score JIF of $1.89\pm0.05$ versus $1.45\pm0.02$; $p$-value~$<10^{-16}$, permutation test). We find similar results for the SJR data.

We have also quantified whether outlier researchers have a preference for a particular outlier sector. To do so, we consider only career years in outlier sectors to estimate their corresponding fractions and calculate the normalized Shannon entropy for every outlier researcher outperforming in more than one category. Entropy values close to unity represent more alternating behaviors, while values around zero indicate that these researchers prefer a given outlier sector. Figure~\ref{fig:1}D shows that the distribution of these entropy values has a peak around 0.6 (purple curve), suggesting a preference for particular outlier sectors. However, if we do not consider sector \IPpp{} (the most underpopulated sector), the entropy shifts to higher values and its distribution peaks around 1 (green curve), indicating that there is no preference between sectors \Ipp{} and \Ppp{} for researchers publishing in both sectors. In this aspect, these atypical researchers are not so different from those present only in non-outlier sectors. As shown in Fig.~\ref{fig:1}D (gray curve), non-outlier researchers also do not exhibit a strong preference for any sector over their careers. The same patterns are observed for the SJR data set (Fig.~S4D~\cite{SI}).

Another intriguing question is whether there are more frequent transitions among sectors of the journal prestige versus productivity plane over the researchers' careers. To investigate this hypothesis, we estimate the number of transitions among all plane sectors and compare them with a null model defined by the average number of transitions estimated after randomly shuffling researchers' careers (over 10,000 realizations). This process allows us to estimate the relative excess for all possible transitions (that is, the number of transitions among sectors during consecutive career years minus the average value of this quantity estimated from the shuffled careers further divided by this same average value). Figure~\ref{fig:1}E shows these transition matrices when grouping researchers in outlier and non-outlier categories. Both matrices are almost symmetric and have positive diagonal elements among the highest absolute values, indicating that most transitions have no preferential direction and a short-term trend to remain in the same sector. For outlier researchers, the transitions \IPpp{}$\de$\IPpp{}, \Ipp{}$\de$\Ipp{}, and \Ppp{}$\de$\Ppp{} have the largest excesses among all self-transitions. For non-outliers, \IpPp{}$\de$\IpPp{} and \ImPp{}$\de$\ImPp{} are the self-transitions with largest excesses. Intriguingly, the self-transition \ImPm{}$\de$\ImPm{} (lowest prestige and lowest productivity sector) has an excess that is larger for outlier (23\%) than for non-outlier (7\%) researchers. 

\begin{figure*}[!ht]
  \centering
  \includegraphics[width=1\textwidth, keepaspectratio]{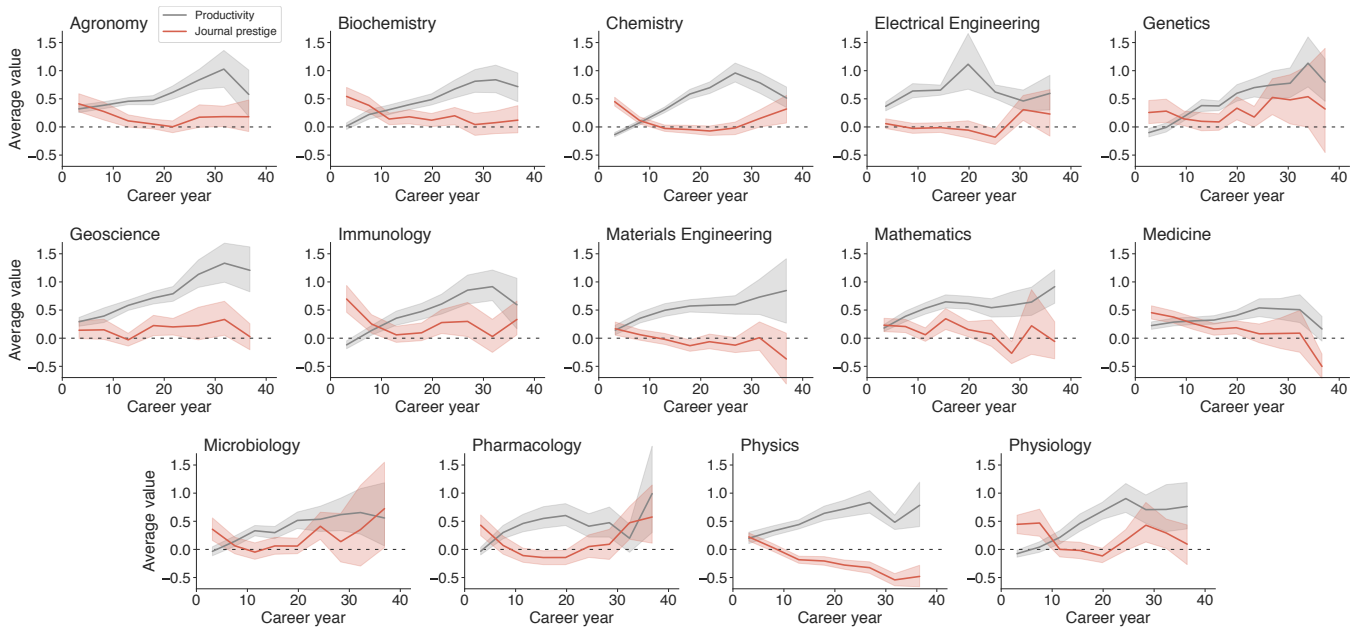}
  \caption{
  \textbf{Average productivity and journal impact over researchers' careers for different disciplines.} These visualizations show the average productivity (gray curves) and the average journal prestige (red curves) calculated within 5-year sliding windows over career years for each discipline in the JIF data set. Shaded areas correspond to bootstrapping 95\% confidence intervals. Average productivity increases with career progression for all disciplines (Fig.~S8A~\cite{SI}) and shows a plateau or small decrease in later career stages for most disciplines. Although some disciplines display more complex patterns, average journal prestige has a subtle downward trend and is usually larger in initial career stages for most disciplines (Fig.~S8B~\cite{SI}).}
  \label{fig:2}
\end{figure*}

The transitions among non-outlier sectors are marked by a negative excess when simultaneously changing levels of productivity and journal impact (\mbox{$I\!\!+\!\!P\pm$}$\ie$\mbox{$I\!\!-\!\!P\mp$}). These transitions represented by the anti-diagonal elements in the non-outlier matrix are less frequent over the careers of outlier and non-outlier researchers. A similar pattern is observed for transitions involving the outlier sectors \Ipp{} and \Ppp{}, that is, the transitions \Ipp{}$\ie$\Ppp{}, \Ppp{}$\de$\IpPm{}, and \Ppp{}$\de$\ImPm{} are also less frequent over the careers of outlier researchers. Conversely, transitions among different sectors with similar productivity or journal prestige (for instance, \IpPp{}$\ie$\ImPp{} and \ImPm{}$\ie$\IpPm{}) usually have excesses close to zero and are thus about as frequent as those occurring in the null model. Together with the excess of self-transitions, these results suggest an aversion to simultaneously changing productivity and journal prestige levels and a preference for maintaining these levels over consecutive years of researchers' careers. 

We further notice that most transitions between outlier and non-outlier sectors occur much less frequently than by chance (negative or close to zero excesses). Career years in sector \Ppp{} are usually not preceded nor followed by years in low productivity sectors (\IpPm{} and \ImPm{}). Conversely, career years in sector \Ipp{} are less followed and less preceded by years in low journal prestige sectors (\ImPp{} and \ImPm{}). It is also worth noticing that career years in the sector \IPpp{} are more often preceded by years in sector \Ppp{} than \Ipp{}, suggesting that it is easier for hyperprolifics to become hyperprolific-perfectionists than for perfectionist researchers. 

We find overall similar results for the SJR data set (Fig.~S4E~\cite{SI}). The main differences emerge for transitions involving sector \IPpp{}. Outside the diagonal, the two largest transitions for outlier researchers are \IPpp{}$\to$\Ipp{} and \IPpp{}$\to$\Ppp{} with 14\% and 12\% excesses, respectively. This result suggests \Ipp{} and \Ppp{} years are more commonly preceded by \IPpp{} years when considering SJR as the measure of journal prestige. In addition, although sector \IPpp{} is still more often preceded by hyperprolific years (\Ppp{}) than by perfectionist ones (\Ipp{}), the difference is not as substantial as it is for the JIF data set. All other transitions display about the same behavior. We also verify that the SJR results are robust when considering only the disciplines present in the JIF data set (Fig.~S7~\cite{SI}).

\subsection*{Effects of career age}

We have also investigated the effects of researchers' career age on the average journal prestige and productivity. To do so, we consider the year after Ph.D. graduation as the first career year. Next, we calculate the average productivity and average journal prestige within 5-year sliding windows of career years for all disciplines. Figure~\ref{fig:2} shows these average values as a function of researchers' career years. We observe a significantly increasing trend of average productivity over the years for all disciplines (Fig.~S8A~\cite{SI}), followed by a plateau or slight decrease in the latest career years. For average journal prestige, while some disciplines show complex patterns, we observe these values are slightly larger during first career years and present a subtle downward trend for most disciplines (Fig.~S8B~\cite{SI}). Figures~S9 and S10~\cite{SI} show similar results for the SJR data set. We remark however that these average trends for disciplines may not represent the individual behavior of researchers, as we shall discuss in the next section.

\begin{figure*}[ht!]
  \centering
  \includegraphics[width=\textwidth]{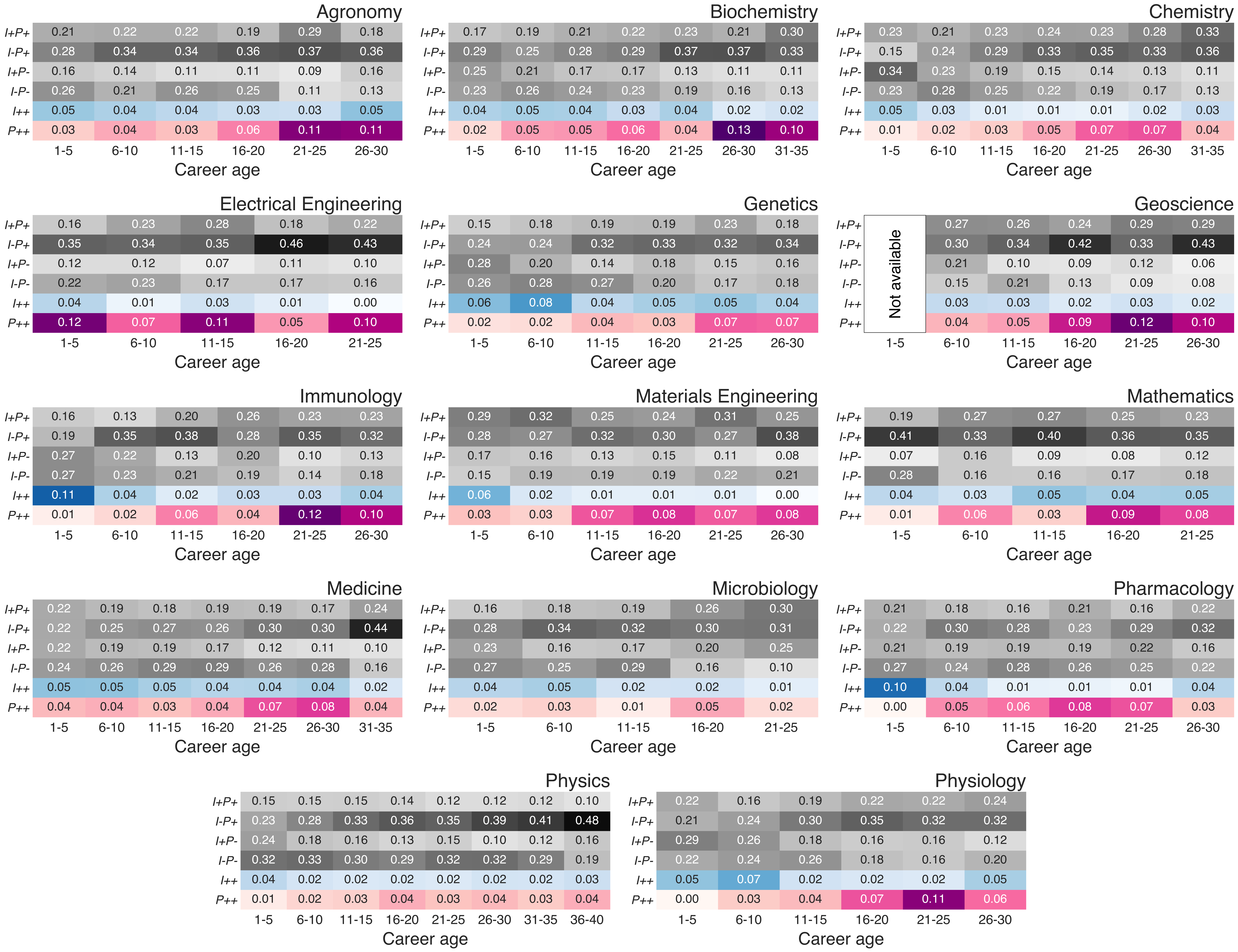}
  \caption{  
  \textbf{Occupation trends in the journal prestige versus productivity plane over researchers' careers.} These visualizations show the fraction of career years in each non-outlier sector and in outlier sectors \Ipp{} and \Ppp{} as a function of researchers' career age for the 14 disciplines in the JIF data set. Columns indicate 5-year intervals and lines represent the different sectors. The same color code indicates the fractions for the non-outlier sectors (gray shades), and the other two color codes are used for the outlier sectors \Ipp{} (blue shades) and \Ppp{} (pink shades). Sector \IPpp{} is omitted because career years in this sector are very rare. We observe that low-productivity sectors are more populated during initial career years and a shifting trend towards high-productivity sectors in later career stages for most disciplines. Only 5-year intervals having at least 20 researchers are shown in these visualizations.
  }
  \label{fig:3}
\end{figure*}

To further characterize the effects of researchers' career age on productivity and journal prestige, we divide scholarly careers into five-year intervals and estimate the average fraction of career years in each sector of the journal prestige versus productivity plane as a career age function. Figure~\ref{fig:3} shows these fractions for all disciplines in our study. In this matrix representation, columns stand for career-years intervals, lines indicate different plane sectors, and color codes stand for the fraction values. Because our data comprise researchers at different career stages, this analysis spans a time interval in career years larger than the number of years in the JIF data set (19 years). 

Figure~\ref{fig:3} indicates that occupation trends in the journal prestige versus productivity plane vary among disciplines (see Fig.~S11~\cite{SI} for results based on the SJR data set). However, some evolution patterns are common. By analyzing the non-outlier sectors, we observe a concentration in low-productivity sectors (\IpPm{} and \ImPm{}) during initial career years and a shifting trend to high-productivity sectors (\IpPp{} and \ImPp{}) in later career stages of researchers from most disciplines. This trend is particularly evident in Physics and Chemistry, for which we observe a more pronounced growth in sector \ImPp{}. For the outlier sectors, we notice a low prevalence in sector \Ppp{} during initial career stages and an increasing trend over time for all disciplines. This rise in productivity levels over the years may reflect the consolidation of researchers' careers and the likely increase of their scientific collaborations. Furthermore, these patterns for non-outlier and outlier researchers agree with the overall increasing trend in average productivity for all disciplines observed in Fig.~\ref{fig:2}. 

Conversely, it is intriguing to observe that sector \Ipp{} tends to be more populated during the initial stages of researchers' careers -- a result that partially explains the slightly larger average journal prestige during first career years for most disciplines observed in Fig.~\ref{fig:2}. This behavior not only indicates that it is more likely to become an impact outlier in initial career years, but also that younger researchers (those having shorter career paths) may outperform more often in this category. Indeed, among the outlier researchers, the chance of finding perfectionist researchers decreases from 79\% to 58\% when career length increases from 10 to 30 years (Fig.~S12A~\cite{SI}). It is worth mentioning this trend of exhibiting high journal prestige in initial careers stages may also reflect a selection effect as our data set only includes researchers belonging to the scientific elite of Brazil. Results for the SJR data set corroborate this finding (Fig.~S12B~\cite{SI}) and indicate very similar trends not only for disciplines present in both data sets but also for disciplines exclusive of the SJR data set.

\subsection*{Quantifying the effect of productivity on journal prestige}

While our findings indicate a negative association between productivity and journal prestige at very high levels of both quantities for most researchers, we have not yet explored this relationship for researchers who never accessed outlier sectors. These non-outlier academics represent 70\% of the researchers in our data set and may exhibit heterogeneous behaviors, limiting the emergence of a clear aggregated relationship at discipline level. To account for these individual behaviors, we select only the productive years of non-outlier researchers with careers longer than five years (see Tables~S1 and~S2~\cite{SI} for details regarding this data set) and use a hierarchical Bayesian model (see Methods for details) for probing the association between productivity and average journal prestige. We assume a linear relationship between journal prestige and productivity, where the distribution of the linear coefficient related to each researcher has a mean drawn from another distribution with average value $\mu_P$. 

\begin{figure*}[!ht]
  \centering
  \includegraphics[width=1\textwidth, keepaspectratio]{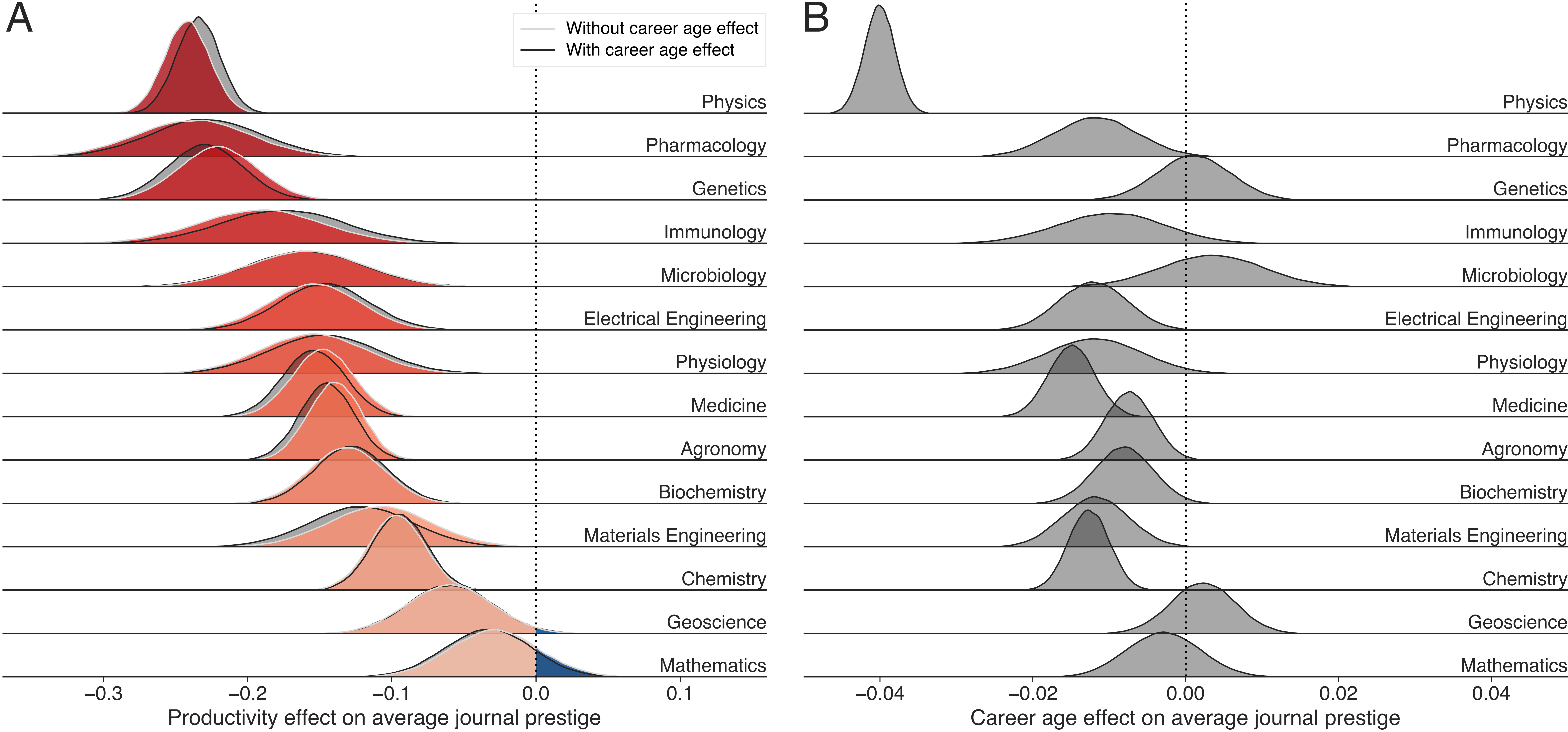}
  \caption{
  \textbf{Effect of productivity on journal prestige for non-outlier researchers.} (A) Posterior probability distributions of the average value of the linear coefficient ($\mu_P$) when considering the association between productivity and journal impact for non-outlier researchers of each discipline. The colored-filled curves represent the results without accounting for the effects of career age, while the gray-filled curves show the distributions of $\mu_P$ after including career age as a confounding factor in the hierarchical Bayesian model (see Methods for details). (B) Posterior probability distributions of the average value of the linear coefficient ($\mu_A$) related to the effect of career age on journal impact for non-outlier researchers of each discipline.
  }
  \label{fig:4}
\end{figure*}

By fitting this model to data with the Bayesian approach, we estimate the posterior probability distribution of the linear coefficient of each researcher and the posterior distribution of $\mu_P$ for each area. Thus, the distribution of $\mu_P$ represents the aggregated effect of productivity on journal impact for non-outlier researchers in each discipline. Distributions of $\mu_P$ shifted toward positive values represent disciplines where most researchers display a positive association between productivity and journal impact. In contrast, distributions more concentrated in negative values characterize disciplines where an increase in productivity correlates with a decline in journal impact for most researchers.

Figure~\ref{fig:4}A shows that the distribution of $\mu_P$ (colored-filled curves) varies significantly among disciplines. All disciplines but Mathematics have distributions entirely located in values of $\mu_P$ lower than zero, suggesting an overall negative association between productivity and average journal impact for most non-outlier researchers. In the most extreme case, a rise in one unit in the productivity of physicists associates with $\approx$0.242 decrease in average journal impact of their publications (in $z$-score units). On the other extreme, we have Mathematics with distribution located near zero. This result indicates that productivity usually plays a small role on journal impact for most mathematicians, while some may display more intense associations (positive or negative). 

The results of Figs.~\ref{fig:2} and \ref{fig:3} have already shown that career age affects the average productivity and journal prestige when aggregating researchers by their respective disciplines. Thus, we can also expect career age to affect the association between journal prestige and productivity at individual level. This is a critical aspect as the overall negative association reported in Fig.~\ref{fig:4}A may reflect a change from an early-career stage marked by low productivity and high impact to higher productivity and lower impact over the years. 

To account for the possible confounding effect of career age on the association between journal prestige and productivity, we have included career age as a predictor of journal impact in the linear hierarchical Bayesian model. In this case, the distribution of the linear coefficient related to the effect of career age for each researcher has a mean drawn from another distribution with average value $\mu_A$ (see Methods for details). Figure~\ref{fig:4}B shows that the distributions of $\mu_A$ also vary among disciplines with most having negative or close to zero average values. These results indicate a reduction in the average journal impact over career years for most researchers from most disciplines. While it is hard to directly compare the effects of changing productivity with the effects of career progression, a 10-year career progression has more effect on journal prestige than increasing one unit of productivity ($z$-score) of a typical researcher only for Chemistry and Physics (Fig.~S13~\cite{SI}). Most importantly, Fig.~\ref{fig:4}A shows that the distributions of $\mu_{P}$ with (colored-filled curves) and without (gray-filled curves) the career age effect change very little. Thus, the confounding effect of career age on the overall negative association between journal prestige and productivity is almost negligible -- that is, an increase in productivity associates with a decrease in journal prestige regardless of career age.

The SJR data set (Figs.~S14 and S15~\cite{SI}) extends this analysis for more disciplines and yields similar results for disciplines present in both data sets.

\section*{Discussion}

We have investigated the association between yearly scientific productivity and average journal impact for over six thousand top Brazilian researchers. Our results explore this association across disciplines, career stage and distinguish researchers with outlier and non-outlier performances. Unlike previous works on the subject, our findings explicitly account for temporal inflation of the bibliometric indicators, scale-dependent effect on average journal prestige, and discipline-specific publication practices via robust standard score measures. These procedures have allowed us to build the journal prestige versus productivity plane -- a coherent and straightforward aggregated representation of researchers' performances in productivity and journal impact. From this representation, we have categorized researchers into outliers and non-outliers and further divided outliers into three categories: hyperprolific (outlier only in productivity), perfectionist (outlier only in journal impact), and hyperprolific-perfectionist (simultaneously outlier in productivity and journal impact).

Researchers with outlier performance comprise 30\% of total scholars in our data set, and the most common behavior is performing as an outlier in only one career year (47.6\% of cases). Among the outliers, the vast majority of researchers are exclusively hyperprolific or exclusively perfectionist. Despite that, 16 extremely hyperprolific researchers display career years solely in sector \IPpp{} when performing above a productivity threshold of $P>27.7$. Only 14.4\% of outlier researchers manage to be hyperprolific and perfectionist over their careers, and solely 6.7\% simultaneously outperform in both categories (the hyperprolific-perfectionists). This former group of 14.4\% of outlier researchers (261 individuals) does not have a preferential outlier sector, displays productivity levels higher than exclusively hyperprolific and perfectionist scholars, and publishes in journals of higher prestige in comparison with exclusively hyperprolific and perfectionist researchers. Furthermore, we find that an increase in the number of hyperprolific career years reduces the probability of performing as a perfectionist for researchers who non-simultaneously outperform in both categories for all disciplines except Materials Engineering in our data set. This negative association varies among disciplines, with Mathematics presenting the most negative effect and Physics the blandest effect. Together, these findings corroborate a negative association between productivity and journal prestige at outlier levels of both quantities. It is extremely hard for researchers to maintain overly high productivity while frequently publishing in very prestigious journals.

We have also explored short-term career patterns regarding productivity and journal impact. To do so, we have estimated the excess of transitions among sectors of the journal prestige versus productivity plane during consecutive career years of outlier and non-outlier researchers. We have identified a persistent behavior in which researchers tend to stay within the same sector of the plane and thus display similar performance over consecutive years. Transitions among similar levels of productivity and journal prestige are about as frequent as by chance. Conversely, transitions among plane sectors with different productivity and journal impact levels occur much less often than by chance, indicating that researchers are averse to simultaneously changing their productivity and journal impact levels over consecutive career years. 

We believe this aversion to simultaneously changing productivity and journal impact and the persistence in maintaining similar performances regarding these metrics indicate possible research and publication strategies in which researchers opt between productivity-focused or journal-impact-focused strategies~\cite{kolesnikov2018researchers}. To keep productivity levels, scholars may choose strategies based on expanding collaborations, avoiding prestigious journals, producing bite-sized articles, and selecting more traditional research themes. Conversely, impact-focused strategies may rely on searching for collaborators only when necessary and beneficial to solving research tasks, selecting high-impact journals as the top-choice, publishing findings with maximization of understanding and impact in mind, and choosing novel research fields. Further research is needed for explicitly identifying these strategies. Still, our results suggest that publication strategies may persist as a habit, and they possibly reflect individual characteristics and cultural conventions of research groups.

We have investigated the aggregated effect of career age on the average journal prestige and productivity for all disciplines. We have identified that journal prestige is usually slightly larger in initial career stages with a subtle downward trend over career years for most disciplines. Productivity, in its turn, tends to increase over career years for all disciplines. We have also studied the effect of career age on the occupation of sectors of the journal prestige versus productivity plane for each discipline. Our findings indicate that disciplines have distinct occupation fractions of these sectors, reflecting the different publication practices of different fields. However, we have found low productivity sectors (\ImPm{} or \IpPm{}) to be more populated during initial stages of researchers' careers from all disciplines. We have also identified an increasing occupation trend of high productivity sectors -- including the hyperprolific sector (\Ppp{}) -- in later career stages for practically all disciplines. Conversely, researchers more often achieve perfectionist performances in early career stages. It is important to remark that both the trend of presenting larger journal prestige in initial career years and the higher probability of finding researchers occupying sector \Ipp{} in yearly career stages may reflect a selection effect, as all researchers in our data set belong to the Brazilian scientific elite. Whether these trends would also hold for other academics is an interesting question that future research could address. The increase in productivity with career age was also verified by Sinatra \textit{et al.}~\cite{sinatra2016quantifying} and may reflect a series of achievements that tend to be usual in scientific career progression, such as more familiarity with research themes~\cite{lawani1986some}, larger availability of financial resources~\cite{lawani1986some,hicks2012performance}, and invitation to write review articles~\cite{lawani1986some}. Similarly, the emergence of hyperprolific years in later career stages may coincide with achieving higher positions in research centers, which could overly enhance publication rates by the tradition of some research disciplines (such as in medical and life sciences) of including the head of scientific labs in all publications~\cite{price2018seventy}. 

Our results have also shown that the relation between productivity and journal impact for non-outlier researchers is similar to the one observed for those achieving outlier performance. For non-outlier, we have used a Bayesian hierarchical model that accounts for researchers' heterogeneous behaviors and identifies the emergent pattern for each discipline. We have found an overall negative association for the majority of disciplines when considering only non-outlier researchers -- a result that is in line with the negative association observed at outlier levels of productivity. However, the intensity of the association varies among disciplines, with Physics having the most negative association and Mathematics having the blandest effect of productivity on journal prestige. We have verified that while career age is also negatively correlated with journal impact, the overall negative association between journal impact and productivity is not significantly affected by this confounding factor. These findings somehow contradict the Nijstad \textit{et al.} ``dual pathway to creativity model''~\cite{nijstad2010dual}, which states that creativity -- as perceived as novel and suitable ideas -- can be achieved through flexibility (usage of a variety of ideas to generate new ones) and persistence (exploration of the same subject in depth) pathways. According to this theory, researchers with high productivity should be either exploring and associating various themes, enabling the generation of creative ideas in the flexibility pathway or intensively working and publishing on the same theme until creative ideas are generated in the persistence pathway. In this sense, since productivity does not positively correlate with journal prestige, the JIF and SJR may not be the most suitable indicators for evaluating creativity.

\subsection*{Data}\label{sec:data}

The data used in our study were obtained from the Lattes Platform~\footnote{\url{http://lattes.cnpq.br/}} (\textit{Plataforma Lattes}). This information system has been maintained by the Brazilian government since 1999 and hosts the official curricula vitae (CV) of academics in Brazil. The Lattes CV is widely used for individual and institutional evaluations, and researchers are required to keep their records up to date. We have initially selected all 14,487 Brazilian researchers (from 88 disciplines) holding the CNPq Research Productivity Fellowship (as of May 2017) and obtained their complete publication records (1,121,652 articles). We filter out researchers whose CVs were not updated from 1 Jan 2016 and those having no information about discipline and Ph.D. conclusion date, reducing the number to 14,146 researchers. We further fill in missing information about publication year and journal by using the DOI reference with the CrossRef API. 

To define the journal prestige of these publications, we have obtained the Journal Impact Factor (JIF) for all available scientific journals between 1997 and 2015 from Clarivate's Journal Citation Reports. We thus combine these two data sets to assign the time-varying values of JIF to the articles published by the CNPq fellows. For each of these researchers, we calculate the number of articles published by year (productivity) and the average JIF of these publications (average journal prestige). Finally, we group these time series by discipline and select the 14 disciplines having at least 50 researchers with published articles in each year between 1997 and 2015. This process leads to our final data set comprising 6,028 researchers from 14 disciplines and 312,881 articles (Fig.~S16~\cite{SI}). We have also considered the Scopus' SCImago Journal Rank (SJR) as a measure of journal prestige. To do so, we have obtained the SJR values for all available scientific journals between 1999 to 2015 from Scopus. By following the same approach used for the JIF, we obtain the productivity and the average SJR for 448,959 articles published by 8,465 researchers from 25 disciplines (Fig.~S17~\cite{SI}).

While the JIF of a journal is simply defined as the number of citations received by articles from the two preceding years divided by the number of published articles in these two previous years~\cite{garfield1972citation, garfield2006if}, the SJR is a more complex network-based (an eigenfactor variant of the PageRank algorithm) indicator~\cite{guerrerobote2012sjr2}. Despite this difference, JIF and SJR are strongly correlated with Pearson correlation $\approx$0.85 (Fig.~S1~\cite{SI}). The disciplines from both data sets are from science, technology, engineering, and mathematics (STEM disciplines), which in turn reflects the predominance of productivity fellowship grants to researchers from these academic disciplines. 

\subsection*{Inflation and robust standard score units}

The volume of scientific production has increased over time at global and individual levels~\cite{solla1963little,sinatra2016quantifying}. This yields an inflation effect that prevents a direct comparison of raw productivity and journal impact values from different periods (Figs.~S2 and S3~\cite{SI}). Disciplines also have distinct volumes of publication and citation dynamics~\cite{bordons2002advantages,foster2015tradition}, which in turn hampers the aggregation and comparison of raw productivity and journal impact values among different disciplines. Furthermore, average journal impact suffers from an additional size effect that decreases its variability with the rise of productivity. This size effect has been observed when comparing the impact factor of journals with different numbers of total publications~\cite{antonoyiannakis2018impact,antonoyiannakis2020impact} and, as argued by Antonoyiannakis~\cite{antonoyiannakis2018impact,antonoyiannakis2020impact}, it represents a direct consequence of the Central Limit Theorem.

To account for these issues, we have used $z$-score measures relative to year and discipline for productivity and $z$-score measures relative to year, discipline, and productivity level for journal prestige. Let $p_{j}^{k}(y)$ and $i_{j}^{k}(y)$ represent, respectively, the number of papers and the average journal prestige of the publications by researcher $j$ from discipline $k$ in year $y$. We calculate the $z$-scores of productivity as
\begin{equation*}
  P_{j}^{k}(y) =  \frac{p_{j}^{k}(y) - {\mathbb{E}[p_{j}^{k}(y)]}}{{\mathbb{S}[p_{j}^{k}(y)]}}\,,
\end{equation*}
where $\mathbb{E}[p_{j}^{k}(y)]$ and $\mathbb{S}[p_{j}^{k}(y)]$ are (respectively) the average and the standard deviation of the productivity of researchers from discipline $k$ in year $y$. Similarly, we calculate the $z$-score journal prestige as
\begin{equation*}
  I_{j}^{k}(y) =  \frac{i_{j}^{k}(y) - {\mathbb{E}[i_{\text{rnd}}^{k}(y, p_{j}^{k}(y))]}}{{\mathbb{S}[i_{\text{rnd}}^{k}(y, p_{j}^{k}(y))]}}\,,
\end{equation*}
where $i_{\text{rnd}}^{k}(y,p)$ is the average journal impact of a random sample of $p$ publications from discipline $k$ in year $y$, and $\mathbb{E}[i_{\text{rnd}}^{k}(y,p)]$ and $\mathbb{S}[i_{\text{rnd}}^{k}(y,p)]$ represent, respectively, the average and the standard deviation of $i_{\text{rnd}}^{k}(y,p)$ estimated over 1,000 independent realizations. This definition is an adaptation of the $\Phi$ index proposed by Antonoyiannakis~\cite{antonoyiannakis2018impact,antonoyiannakis2020impact} for ranking journals of different sizes, and it accounts for the fact that low-productivity researchers have high variability in their average values of journal prestige, while high-productivity researchers display significantly lower variability (Figs.~S18~\cite{SI}). 

Here we have further used Huber robust estimators for mean (location) and standard deviation (scale) in place of the usual estimators~\cite{huber2004robust} due to the existence of outlier values for $p_{j}^{k}(y)$ and $i_{j}^{k}(y)$ (Fig.~S19~\cite{SI}) -- that is, $\mathbb{E}[\dots]$ and $\mathbb{S}[\dots]$ represent respectively Huber's estimators of location and scale (as implemented in Python package \textit{statsmodels}~\cite{statsmodels}).

\subsection*{Logistic regressions}

To quantify the effect of performing as outlier in productivity on the chance of being an outlier in journal impact (perfectionist), we use the following logistic model
\begin{equation*}
  \Pi_{\text{perfectionist}} = \frac{e^{\alpha_0 + \alpha_1 Y_P}}{1-e^{\alpha_0 + \alpha_1 Y_P}}\,,
\end{equation*}
where $\Pi_{\text{perfectionist}}$ is the probability of being a perfectionist researcher given the scholar has $Y_P$ outlier career years in productivity, $\alpha_0$ is the intercept, and $\alpha_1$ is the logistic regression coefficient. Positive values of $\alpha_1$ indicate that an increase in $Y_P$ enhances the probability of performing as a perfectionist, while negative values of $\alpha_1$ show that an increase in $Y_P$ reduces the probability of being a perfectionist. We have adjusted this model (as implemented in the Python package \textit{statsmodels}~\cite{statsmodels}) to our data by considering all researchers who outperformed in journal impact or productivity at some point in their careers. We have also adjusted the same model when grouping researchers by discipline. Figure~\ref{fig:1}C shows $\Pi_{\text{perfectionist}}$ as a function of $Y_P$ for each discipline in the JIF data set and when considering all disciplines together (the inset depicts the values of $\alpha_1$). Similarly, Fig.~S4C~\cite{SI} shows the corresponding results for the SJR data set.

We further use a similar logistic model to estimate the probability of being perfectionist as a function of the career length of researchers ($L$). In this case, the model can be written as
\begin{equation*}
  \Pi_{\text{perfectionist}} = \frac{e^{\theta_0 + \theta_1 L}}{1-e^{\theta_0 + \theta_1 L}}\,,
\end{equation*}
where $\theta_0$ (intercept) and $\theta_1$ (regression coefficient) are the model parameters. We have adjusted this model considering all outlier researchers in JIF and SJR data sets. Figure~S12~\cite{SI} shows $\Pi_{\text{perfectionist}}$ as a function of $L$ for both data sets. The adjusted parameters are $\theta_0 = 1.849 \pm 0.132$ and  $\theta_1 = -0.051 \pm 0.006$ for the JIF data set, and $\theta_0 = 1.921 \pm 0.108$ and $\theta_1 = -0.054 \pm 0.005$ for the SJR data set.

\subsection*{Bayesian hierarchical model}
We use a Bayesian hierarchical model to estimate the effect of productivity on average journal impact for non-outlier researchers. For a given discipline, we consider that the data are hierarchically structured such that each observation of $I_j$ and $P_j$ is nested within a researcher $j$ (here we have dropped the index $k$ for simplicity). We further assume a linear relation between these variables at the individual level, where $c_j$ and $\beta_{j}$ are (respectively) the intercept and the slope of the linear association for the $j$-th researcher of a given discipline. We consider the parameters $c_j$ and $\beta_{j}$ as random variables distributed according to normal distributions whose parameters are also random variables. Mathematically, we can write this model as
\begin{equation}\label{eq:lmm}
  I_j \sim \mathcal{N}(c_j + \beta_{j} P_j, \varepsilon)\,,
\end{equation}
where $\mathcal{N}(\mu,\sigma)$ stands for a normal distribution with mean $\mu$ and standard deviation $\sigma$, $\varepsilon$ accounts for the unobserved determinants of $I_j$, and
\begin{equation*}
\begin{split}
  c_j     & \sim \mathcal{N}(\mu_c, \sigma_c) \\
  \beta_j & \sim \mathcal{N}(\mu_P, \sigma_P) \\
\end{split}\,,
\end{equation*}
where $\mu_c$ is the mean and $\sigma_c$ is the standard deviation of a normal distribution associated with the intercept $c_j$, and $\mu_P$ and $\sigma_P$ are the same for the distribution associated with $\beta_j$. The Bayesian inference process consists in determining the posterior probability distributions of the parameters at discipline level ($\mu_c$, $\sigma_c$, $\mu_P$ and $\sigma_P$) and at researcher level ($c_j$  and $\beta_{j}$ for every $j$ researcher of given discipline).

We perform this Bayesian regression for each area separately and use non-informative prior distributions~\cite{Gelman06priordistributions} to not bias the posterior estimation, that is, we consider
\begin{equation}\label{eq:priors}
\begin{split}
  \varepsilon & \sim \mathcal{U}(0,10^2) \\
  \mu_c       & \sim \mathcal{N}(0, 10^5) \\
  \mu_P       & \sim \mathcal{N}(0, 10^5) \\
  \sigma_c    & \sim \text{Inv-}\Gamma(10^{-3}, 1) \\
  \sigma_P    & \sim \text{Inv-}\Gamma(10^{-3}, 1) \\
\end{split}\,,
\end{equation}
where $\mathcal{U}(x_{\text{min}},x_{\text{max}})$ represents a uniform distribution between $x_{\text{min}}$ and $x_{\text{max}}$, and $\text{Inv-}\Gamma(a,b)$ stands for an inverse-gamma distribution with parameters $a$ (shape) and $b$ (scale). Figure~S20~\cite{SI} shows a graphical representation of this model. 

We have also considered a generalized version of the model defined in Eq.~\ref{eq:lmm}, where career age $A_j$ is also assumed to be linearly related with average journal prestige. The value of $A_j$ refers to the career age of researcher $j$ at a given year $y$ with productivity $P_j$ and average journal impact $I_j$. Thus, we include the career age $A_j$ as an independent variable in the hierarchical model of Eq.~\ref{eq:lmm}, yielding
\begin{equation}\label{eq:glmm}
  I_j \sim \mathcal{N}(c_j + \beta_{j} P_j + \gamma_{j} A_j, \varepsilon)\,,
\end{equation}
where $\gamma_{j}$ is the slope of the linear association between career age and journal prestige. This linear coefficient is assumed to be distributed according to a normal distribution
\begin{equation*}
  \gamma_j \sim \mathcal{N}(\mu_A, \sigma_A)\,,
\end{equation*}
where $\mu_A$ is the mean and $\sigma_A$ is the standard deviation. We adjust the model of Eq.~\ref{eq:glmm} with the same non-informative prior distributions defined in Eq.~\ref{eq:priors}, and use 
\begin{equation}
\begin{split}
  \mu_A       & \sim \mathcal{N}(0, 10^5) \\
  \sigma_A    & \sim \text{Inv-}\Gamma(10^{-3}, 1) \\
\end{split}\,
\end{equation}
as the non-informative prior distributions for the additional parameters related to the effects of career age. Figure~S21~\cite{SI} shows a graphical representation of this generalized model that accounts for possible confounding effects of career age on the association between average journal prestige and productivity.

We implement these two models (Eqs.~\ref{eq:lmm} and \ref{eq:glmm}) using the PyMC3 framework~\cite{pymc3} via gradient-based Hamiltonian Monte Carlo No-U-Turn-Sampler method for sampling the posterior distributions. We run $8$ parallel chains with 10,000 iterations (of which 5,000 are burn-in samples) to allow well-mixing of the Monte Carlo chains. We estimate the Gelman-Rubin convergence statistic (R-hat) for all regression analyses and the results were all close to one, an indication of convergence of the sampling approach. 

\begin{acknowledgments}
A.S.S. and H.V.R acknowledge the support of the Coordena\c{c}\~ao de Aperfei\c{c}oamento de Pessoal de N\'ivel Superior (CAPES) and the Conselho Nacional de Desenvolvimento Cient\'ifico e Tecnol\'ogico (CNPq -- Grants 407690/2018-2 and 303121/2018-1). M.P. acknowledges the support of the Slovenian Research Agency (Grants J1-2457 and P1-0403).
\end{acknowledgments}

\bibliography{prr_refs}

\begin{thebibliography}{61}%
\makeatletter
\providecommand \@ifxundefined [1]{%
 \@ifx{#1\undefined}
}%
\providecommand \@ifnum [1]{%
 \ifnum #1\expandafter \@firstoftwo
 \else \expandafter \@secondoftwo
 \fi
}%
\providecommand \@ifx [1]{%
 \ifx #1\expandafter \@firstoftwo
 \else \expandafter \@secondoftwo
 \fi
}%
\providecommand \natexlab [1]{#1}%
\providecommand \enquote  [1]{``#1''}%
\providecommand \bibnamefont  [1]{#1}%
\providecommand \bibfnamefont [1]{#1}%
\providecommand \citenamefont [1]{#1}%
\providecommand \href@noop [0]{\@secondoftwo}%
\providecommand \href [0]{\begingroup \@sanitize@url \@href}%
\providecommand \@href[1]{\@@startlink{#1}\@@href}%
\providecommand \@@href[1]{\endgroup#1\@@endlink}%
\providecommand \@sanitize@url [0]{\catcode `\\12\catcode `\$12\catcode
  `\&12\catcode `\#12\catcode `\^12\catcode `\_12\catcode `\%12\relax}%
\providecommand \@@startlink[1]{}%
\providecommand \@@endlink[0]{}%
\providecommand \url  [0]{\begingroup\@sanitize@url \@url }%
\providecommand \@url [1]{\endgroup\@href {#1}{\urlprefix }}%
\providecommand \urlprefix  [0]{URL }%
\providecommand \Eprint [0]{\href }%
\providecommand \doibase [0]{https://doi.org/}%
\providecommand \selectlanguage [0]{\@gobble}%
\providecommand \bibinfo  [0]{\@secondoftwo}%
\providecommand \bibfield  [0]{\@secondoftwo}%
\providecommand \translation [1]{[#1]}%
\providecommand \BibitemOpen [0]{}%
\providecommand \bibitemStop [0]{}%
\providecommand \bibitemNoStop [0]{.\EOS\space}%
\providecommand \EOS [0]{\spacefactor3000\relax}%
\providecommand \BibitemShut  [1]{\csname bibitem#1\endcsname}%
\let\auto@bib@innerbib\@empty
\bibitem [{\citenamefont {Zeng}\ \emph {et~al.}(2017)\citenamefont {Zeng},
  \citenamefont {Shen}, \citenamefont {Zhou}, \citenamefont {Wu}, \citenamefont
  {Fan}, \citenamefont {Wang},\ and\ \citenamefont
  {Stanley}}]{zeng2017science}%
  \BibitemOpen
  \bibfield  {author} {\bibinfo {author} {\bibfnamefont {A.}~\bibnamefont
  {Zeng}}, \bibinfo {author} {\bibfnamefont {Z.}~\bibnamefont {Shen}}, \bibinfo
  {author} {\bibfnamefont {J.}~\bibnamefont {Zhou}}, \bibinfo {author}
  {\bibfnamefont {J.}~\bibnamefont {Wu}}, \bibinfo {author} {\bibfnamefont
  {Y.}~\bibnamefont {Fan}}, \bibinfo {author} {\bibfnamefont {Y.}~\bibnamefont
  {Wang}},\ and\ \bibinfo {author} {\bibfnamefont {H.~E.}\ \bibnamefont
  {Stanley}},\ }\bibfield  {title} {\bibinfo {title} {The science of science:
  {From} the perspective of complex systems},\ }\href
  {https://doi.org/10.1016/j.physrep.2017.10.001} {\bibfield  {journal}
  {\bibinfo  {journal} {Physics Reports}\ }\textbf {\bibinfo {volume} {714}},\
  \bibinfo {pages} {1} (\bibinfo {year} {2017})}\BibitemShut {NoStop}%
\bibitem [{\citenamefont {Fortunato}\ \emph {et~al.}(2018)\citenamefont
  {Fortunato}, \citenamefont {Bergstrom}, \citenamefont {B{\"o}rner},
  \citenamefont {Evans}, \citenamefont {Helbing}, \citenamefont
  {Milojevi{\'c}}, \citenamefont {Petersen}, \citenamefont {Radicchi},
  \citenamefont {Sinatra}, \citenamefont {Uzzi}, \citenamefont {Vespignani},
  \citenamefont {Waltman}, \citenamefont {Wang},\ and\ \citenamefont
  {Barab{\'a}si}}]{fortunato2018science}%
  \BibitemOpen
  \bibfield  {author} {\bibinfo {author} {\bibfnamefont {S.}~\bibnamefont
  {Fortunato}}, \bibinfo {author} {\bibfnamefont {C.~T.}\ \bibnamefont
  {Bergstrom}}, \bibinfo {author} {\bibfnamefont {K.}~\bibnamefont
  {B{\"o}rner}}, \bibinfo {author} {\bibfnamefont {J.~A.}\ \bibnamefont
  {Evans}}, \bibinfo {author} {\bibfnamefont {D.}~\bibnamefont {Helbing}},
  \bibinfo {author} {\bibfnamefont {S.}~\bibnamefont {Milojevi{\'c}}}, \bibinfo
  {author} {\bibfnamefont {A.~M.}\ \bibnamefont {Petersen}}, \bibinfo {author}
  {\bibfnamefont {F.}~\bibnamefont {Radicchi}}, \bibinfo {author}
  {\bibfnamefont {R.}~\bibnamefont {Sinatra}}, \bibinfo {author} {\bibfnamefont
  {B.}~\bibnamefont {Uzzi}}, \bibinfo {author} {\bibfnamefont {A.}~\bibnamefont
  {Vespignani}}, \bibinfo {author} {\bibfnamefont {L.}~\bibnamefont {Waltman}},
  \bibinfo {author} {\bibfnamefont {D.}~\bibnamefont {Wang}},\ and\ \bibinfo
  {author} {\bibfnamefont {A.-L.}\ \bibnamefont {Barab{\'a}si}},\ }\bibfield
  {title} {\bibinfo {title} {Science of science},\ }\href
  {https://doi.org/10.1126/science.aao0185} {\bibfield  {journal} {\bibinfo
  {journal} {Science}\ }\textbf {\bibinfo {volume} {359}},\ \bibinfo {pages}
  {eaao0185} (\bibinfo {year} {2018})}\BibitemShut {NoStop}%
\bibitem [{\citenamefont {Azoulay}\ \emph {et~al.}(2011)\citenamefont
  {Azoulay}, \citenamefont {Zivin},\ and\ \citenamefont
  {Manso}}]{azoulay2009incentives}%
  \BibitemOpen
  \bibfield  {author} {\bibinfo {author} {\bibfnamefont {P.}~\bibnamefont
  {Azoulay}}, \bibinfo {author} {\bibfnamefont {J.~S.~G.}\ \bibnamefont
  {Zivin}},\ and\ \bibinfo {author} {\bibfnamefont {G.}~\bibnamefont {Manso}},\
  }\bibfield  {title} {\bibinfo {title} {Incentives and creativity: {Evidence}
  from the academic life sciences},\ }\href
  {https://doi.org/10.1111/j.1756-2171.2011.00140.x} {\bibfield  {journal}
  {\bibinfo  {journal} {The RAND Journal of Economics}\ }\textbf {\bibinfo
  {volume} {42}},\ \bibinfo {pages} {527} (\bibinfo {year} {2011})}\BibitemShut
  {NoStop}%
\bibitem [{\citenamefont {Bromham}\ \emph {et~al.}(2016)\citenamefont
  {Bromham}, \citenamefont {Dinnage},\ and\ \citenamefont
  {Hua}}]{hua2016interdisciplinary}%
  \BibitemOpen
  \bibfield  {author} {\bibinfo {author} {\bibfnamefont {L.}~\bibnamefont
  {Bromham}}, \bibinfo {author} {\bibfnamefont {R.}~\bibnamefont {Dinnage}},\
  and\ \bibinfo {author} {\bibfnamefont {X.}~\bibnamefont {Hua}},\ }\bibfield
  {title} {\bibinfo {title} {Interdisciplinary research has consistently lower
  funding success},\ }\href {https://doi.org/10.1038/nature18315} {\bibfield
  {journal} {\bibinfo  {journal} {Nature}\ }\textbf {\bibinfo {volume} {534}},\
  \bibinfo {pages} {684} (\bibinfo {year} {2016})}\BibitemShut {NoStop}%
\bibitem [{\citenamefont {Meirmans}\ \emph {et~al.}(2019)\citenamefont
  {Meirmans}, \citenamefont {Butlin}, \citenamefont {Charmantier},
  \citenamefont {Engelst{\"a}dter}, \citenamefont {Groot}, \citenamefont
  {King}, \citenamefont {Kokko}, \citenamefont {Reid},\ and\ \citenamefont
  {Neiman}}]{meirmans2019science}%
  \BibitemOpen
  \bibfield  {author} {\bibinfo {author} {\bibfnamefont {S.}~\bibnamefont
  {Meirmans}}, \bibinfo {author} {\bibfnamefont {R.~K.}\ \bibnamefont
  {Butlin}}, \bibinfo {author} {\bibfnamefont {A.}~\bibnamefont {Charmantier}},
  \bibinfo {author} {\bibfnamefont {J.}~\bibnamefont {Engelst{\"a}dter}},
  \bibinfo {author} {\bibfnamefont {A.~T.}\ \bibnamefont {Groot}}, \bibinfo
  {author} {\bibfnamefont {K.~C.}\ \bibnamefont {King}}, \bibinfo {author}
  {\bibfnamefont {H.}~\bibnamefont {Kokko}}, \bibinfo {author} {\bibfnamefont
  {J.~M.}\ \bibnamefont {Reid}},\ and\ \bibinfo {author} {\bibfnamefont
  {M.}~\bibnamefont {Neiman}},\ }\bibfield  {title} {\bibinfo {title} {Science
  policies: {How} should science funding be allocated? {An} evolutionary
  biologists’ perspective},\ }\href {https://doi.org/10.1111/jeb.13497}
  {\bibfield  {journal} {\bibinfo  {journal} {Journal of Evolutionary Biology}\
  }\textbf {\bibinfo {volume} {32}},\ \bibinfo {pages} {754} (\bibinfo {year}
  {2019})}\BibitemShut {NoStop}%
\bibitem [{\citenamefont {Wilsdon}(2016)}]{wilsdon2016metric}%
  \BibitemOpen
  \bibfield  {author} {\bibinfo {author} {\bibfnamefont {J.}~\bibnamefont
  {Wilsdon}},\ }\href {https://doi.org/10.4135/9781473978782} {\emph {\bibinfo
  {title} {{The Metric Tide: Independent} review of the role of metrics in
  research assessment and management}}}\ (\bibinfo  {publisher} {Sage},\
  \bibinfo {year} {2016})\BibitemShut {NoStop}%
\bibitem [{\citenamefont {Wessely}(1998)}]{wessely1998peer}%
  \BibitemOpen
  \bibfield  {author} {\bibinfo {author} {\bibfnamefont {S.}~\bibnamefont
  {Wessely}},\ }\bibfield  {title} {\bibinfo {title} {Peer review of grant
  applications: {What} do we know?},\ }\href
  {https://doi.org/10.1016/S0140-6736(97)11129-1} {\bibfield  {journal}
  {\bibinfo  {journal} {The Lancet}\ }\textbf {\bibinfo {volume} {352}},\
  \bibinfo {pages} {301} (\bibinfo {year} {1998})}\BibitemShut {NoStop}%
\bibitem [{\citenamefont {Smith}(2006)}]{smith2006peer}%
  \BibitemOpen
  \bibfield  {author} {\bibinfo {author} {\bibfnamefont {R.}~\bibnamefont
  {Smith}},\ }\bibfield  {title} {\bibinfo {title} {Peer review: {A} flawed
  process at the heart of science and journals},\ }\href
  {https://doi.org/10.1177/014107680609900414} {\bibfield  {journal} {\bibinfo
  {journal} {Journal of the Royal Society of Medicine}\ }\textbf {\bibinfo
  {volume} {99}},\ \bibinfo {pages} {178} (\bibinfo {year} {2006})}\BibitemShut
  {NoStop}%
\bibitem [{\citenamefont {Balietti}\ \emph {et~al.}(2016)\citenamefont
  {Balietti}, \citenamefont {Goldstone},\ and\ \citenamefont
  {Helbing}}]{balietti2016peer}%
  \BibitemOpen
  \bibfield  {author} {\bibinfo {author} {\bibfnamefont {S.}~\bibnamefont
  {Balietti}}, \bibinfo {author} {\bibfnamefont {R.~L.}\ \bibnamefont
  {Goldstone}},\ and\ \bibinfo {author} {\bibfnamefont {D.}~\bibnamefont
  {Helbing}},\ }\bibfield  {title} {\bibinfo {title} {Peer review and
  competition in the art exhibition game},\ }\href
  {https://doi.org/10.1073/pnas.1603723113} {\bibfield  {journal} {\bibinfo
  {journal} {Proceedings of the National Academy of Sciences}\ }\textbf
  {\bibinfo {volume} {113}},\ \bibinfo {pages} {8414} (\bibinfo {year}
  {2016})}\BibitemShut {NoStop}%
\bibitem [{\citenamefont {Bornmann}\ and\ \citenamefont
  {Mutz}(2015)}]{bornmann2015growth}%
  \BibitemOpen
  \bibfield  {author} {\bibinfo {author} {\bibfnamefont {L.}~\bibnamefont
  {Bornmann}}\ and\ \bibinfo {author} {\bibfnamefont {R.}~\bibnamefont
  {Mutz}},\ }\bibfield  {title} {\bibinfo {title} {Growth rates of modern
  science: {A} bibliometric analysis based on the number of publications and
  cited references},\ }\href {https://doi.org/10.1002/asi.23329} {\bibfield
  {journal} {\bibinfo  {journal} {Journal of the Association for Information
  Science and Technology}\ }\textbf {\bibinfo {volume} {66}},\ \bibinfo {pages}
  {2215} (\bibinfo {year} {2015})}\BibitemShut {NoStop}%
\bibitem [{\citenamefont {Ioannidis}\ \emph {et~al.}(2014)\citenamefont
  {Ioannidis}, \citenamefont {Boyack},\ and\ \citenamefont
  {Klavans}}]{ioannidis2014estimates}%
  \BibitemOpen
  \bibfield  {author} {\bibinfo {author} {\bibfnamefont {J.~P.}\ \bibnamefont
  {Ioannidis}}, \bibinfo {author} {\bibfnamefont {K.~W.}\ \bibnamefont
  {Boyack}},\ and\ \bibinfo {author} {\bibfnamefont {R.}~\bibnamefont
  {Klavans}},\ }\bibfield  {title} {\bibinfo {title} {Estimates of the
  continuously publishing core in the scientific workforce},\ }\href
  {https://doi.org/10.1371/journal.pone.0101698} {\bibfield  {journal}
  {\bibinfo  {journal} {PLoS ONE}\ }\textbf {\bibinfo {volume} {9}},\ \bibinfo
  {pages} {e101698} (\bibinfo {year} {2014})}\BibitemShut {NoStop}%
\bibitem [{\citenamefont {Cameron}(2005)}]{cameron2005trends}%
  \BibitemOpen
  \bibfield  {author} {\bibinfo {author} {\bibfnamefont {B.~D.}\ \bibnamefont
  {Cameron}},\ }\bibfield  {title} {\bibinfo {title} {Trends in the usage of
  {ISI} bibliometric data: {Uses}, abuses, and implications},\ }\href
  {https://doi.org/10.1353/pla.2005.0003} {\bibfield  {journal} {\bibinfo
  {journal} {portal: Libraries and the Academy}\ }\textbf {\bibinfo {volume}
  {5}},\ \bibinfo {pages} {105} (\bibinfo {year} {2005})}\BibitemShut {NoStop}%
\bibitem [{\citenamefont {Traag}\ and\ \citenamefont
  {Waltman}(2019)}]{traag2019systematic}%
  \BibitemOpen
  \bibfield  {author} {\bibinfo {author} {\bibfnamefont {V.~A.}\ \bibnamefont
  {Traag}}\ and\ \bibinfo {author} {\bibfnamefont {L.}~\bibnamefont
  {Waltman}},\ }\bibfield  {title} {\bibinfo {title} {{Systematic analysis of
  agreement between metrics and peer review in the UK REF}},\ }\bibfield
  {journal} {\bibinfo  {journal} {Palgrave Communications}\ }\textbf {\bibinfo
  {volume} {5}},\ \href {https://doi.org/10.1057/s41599-019-0233-x}
  {10.1057/s41599-019-0233-x} (\bibinfo {year} {2019})\BibitemShut {NoStop}%
\bibitem [{\citenamefont {Gagolewski}(2013)}]{gagolewski2013scientific}%
  \BibitemOpen
  \bibfield  {author} {\bibinfo {author} {\bibfnamefont {M.}~\bibnamefont
  {Gagolewski}},\ }\bibfield  {title} {\bibinfo {title} {Scientific impact
  assessment cannot be fair},\ }\href
  {https://doi.org/10.1016/j.joi.2013.07.001} {\bibfield  {journal} {\bibinfo
  {journal} {Journal of Informetrics}\ }\textbf {\bibinfo {volume} {7}},\
  \bibinfo {pages} {792} (\bibinfo {year} {2013})}\BibitemShut {NoStop}%
\bibitem [{\citenamefont {Siudem}\ \emph {et~al.}(2020)\citenamefont {Siudem},
  \citenamefont {{\.Z}oga{\l}a-Siudem}, \citenamefont {Cena},\ and\
  \citenamefont {Gagolewski}}]{siudem2020three}%
  \BibitemOpen
  \bibfield  {author} {\bibinfo {author} {\bibfnamefont {G.}~\bibnamefont
  {Siudem}}, \bibinfo {author} {\bibfnamefont {B.}~\bibnamefont
  {{\.Z}oga{\l}a-Siudem}}, \bibinfo {author} {\bibfnamefont {A.}~\bibnamefont
  {Cena}},\ and\ \bibinfo {author} {\bibfnamefont {M.}~\bibnamefont
  {Gagolewski}},\ }\bibfield  {title} {\bibinfo {title} {Three dimensions of
  scientific impact},\ }\href {https://doi.org/10.1073/pnas.2001064117}
  {\bibfield  {journal} {\bibinfo  {journal} {Proceedings of the National
  Academy of Sciences}\ }\textbf {\bibinfo {volume} {117}},\ \bibinfo {pages}
  {13896} (\bibinfo {year} {2020})}\BibitemShut {NoStop}%
\bibitem [{dor(2012)}]{dora}%
  \BibitemOpen
  \href@noop {} {\bibinfo {title} {{San Francisco Declaration on Research
  Assessment}}} (\bibinfo {year} {2012}),\ \bibinfo {note} {available at
  \url{https://sfdora.org/read/}. Accessed August 2020}\BibitemShut {NoStop}%
\bibitem [{\citenamefont {Hicks}\ \emph {et~al.}(2015)\citenamefont {Hicks},
  \citenamefont {Wouters}, \citenamefont {Waltman}, \citenamefont {Rijcke},\
  and\ \citenamefont {Rafols}}]{hicks2015bibliometrics}%
  \BibitemOpen
  \bibfield  {author} {\bibinfo {author} {\bibfnamefont {D.}~\bibnamefont
  {Hicks}}, \bibinfo {author} {\bibfnamefont {P.}~\bibnamefont {Wouters}},
  \bibinfo {author} {\bibfnamefont {L.}~\bibnamefont {Waltman}}, \bibinfo
  {author} {\bibfnamefont {S.~D.}\ \bibnamefont {Rijcke}},\ and\ \bibinfo
  {author} {\bibfnamefont {I.}~\bibnamefont {Rafols}},\ }\bibfield  {title}
  {\bibinfo {title} {Bibliometrics: {The} {Leiden Manifesto} for research
  metrics},\ }\href {https://doi.org/10.1038/520429a} {\bibfield  {journal}
  {\bibinfo  {journal} {Nature}\ }\textbf {\bibinfo {volume} {520}},\ \bibinfo
  {pages} {429} (\bibinfo {year} {2015})}\BibitemShut {NoStop}%
\bibitem [{nuf(2014)}]{nuffield}%
  \BibitemOpen
  \href@noop {} {\bibinfo {title} {{Nuffield Council on Bioethics. The findings
  of a series of engagement activities exploring the culture of scientific
  research in the UK}}} (\bibinfo {year} {2014}),\ \bibinfo {note} {available
  at
  \url{https://www.nuffieldbioethics.org/assets/pdfs/The-culture-of-scientific-research-report.pdf}.
  Accessed September 2020}\BibitemShut {NoStop}%
\bibitem [{\citenamefont {Powell}(2016)}]{powell2016junior}%
  \BibitemOpen
  \bibfield  {author} {\bibinfo {author} {\bibfnamefont {K.}~\bibnamefont
  {Powell}},\ }\bibfield  {title} {\bibinfo {title} {Young, talented and
  fed-up: {Scientists} tell their stories},\ }\href
  {https://doi.org/10.1038/538446a} {\bibfield  {journal} {\bibinfo  {journal}
  {Nature}\ }\textbf {\bibinfo {volume} {538}},\ \bibinfo {pages} {446}
  (\bibinfo {year} {2016})}\BibitemShut {NoStop}%
\bibitem [{\citenamefont {Moher}\ \emph {et~al.}(2018)\citenamefont {Moher},
  \citenamefont {Naudet}, \citenamefont {Cristea}, \citenamefont {Miedema},
  \citenamefont {Ioannidis},\ and\ \citenamefont
  {Goodman}}]{moher2018assessing}%
  \BibitemOpen
  \bibfield  {author} {\bibinfo {author} {\bibfnamefont {D.}~\bibnamefont
  {Moher}}, \bibinfo {author} {\bibfnamefont {F.}~\bibnamefont {Naudet}},
  \bibinfo {author} {\bibfnamefont {I.~A.}\ \bibnamefont {Cristea}}, \bibinfo
  {author} {\bibfnamefont {F.}~\bibnamefont {Miedema}}, \bibinfo {author}
  {\bibfnamefont {J.~P.}\ \bibnamefont {Ioannidis}},\ and\ \bibinfo {author}
  {\bibfnamefont {S.~N.}\ \bibnamefont {Goodman}},\ }\bibfield  {title}
  {\bibinfo {title} {Assessing scientists for hiring, promotion, and tenure},\
  }\href {https://doi.org/10.1371/journal.pbio.2004089} {\bibfield  {journal}
  {\bibinfo  {journal} {PLoS Biology}\ }\textbf {\bibinfo {volume} {16}},\
  \bibinfo {pages} {e2004089} (\bibinfo {year} {2018})}\BibitemShut {NoStop}%
\bibitem [{\citenamefont {Schimanski}\ and\ \citenamefont
  {Alperin}(2018)}]{schimanski2018evaluation}%
  \BibitemOpen
  \bibfield  {author} {\bibinfo {author} {\bibfnamefont {L.~A.}\ \bibnamefont
  {Schimanski}}\ and\ \bibinfo {author} {\bibfnamefont {J.~P.}\ \bibnamefont
  {Alperin}},\ }\bibfield  {title} {\bibinfo {title} {The evaluation of
  scholarship in academic promotion and tenure processes: {Past}, present, and
  future},\ }\href {https://doi.org/10.12688/f1000research.16493.1} {\bibfield
  {journal} {\bibinfo  {journal} {F1000Research}\ }\textbf {\bibinfo {volume}
  {7}},\ \bibinfo {pages} {1} (\bibinfo {year} {2018})}\BibitemShut {NoStop}%
\bibitem [{\citenamefont {Dennis}(1954)}]{dennis1954productivity}%
  \BibitemOpen
  \bibfield  {author} {\bibinfo {author} {\bibfnamefont {W.}~\bibnamefont
  {Dennis}},\ }\bibfield  {title} {\bibinfo {title} {Productivity among
  american psychologists},\ }\href {https://doi.org/10.1037/h0057477}
  {\bibfield  {journal} {\bibinfo  {journal} {American Psychologist}\ }\textbf
  {\bibinfo {volume} {9}},\ \bibinfo {pages} {191} (\bibinfo {year}
  {1954})}\BibitemShut {NoStop}%
\bibitem [{\citenamefont {White}\ and\ \citenamefont
  {White}(1978)}]{white1978relation}%
  \BibitemOpen
  \bibfield  {author} {\bibinfo {author} {\bibfnamefont {K.~G.}\ \bibnamefont
  {White}}\ and\ \bibinfo {author} {\bibfnamefont {M.~J.}\ \bibnamefont
  {White}},\ }\bibfield  {title} {\bibinfo {title} {On the relation between
  productivity and impact},\ }\href {https://doi.org/10.1080/00050067808254325}
  {\bibfield  {journal} {\bibinfo  {journal} {Australian Psychologist}\
  }\textbf {\bibinfo {volume} {13}},\ \bibinfo {pages} {369} (\bibinfo {year}
  {1978})}\BibitemShut {NoStop}%
\bibitem [{\citenamefont {Lawani}(1986)}]{lawani1986some}%
  \BibitemOpen
  \bibfield  {author} {\bibinfo {author} {\bibfnamefont {S.}~\bibnamefont
  {Lawani}},\ }\bibfield  {title} {\bibinfo {title} {Some bibliometric
  correlates of quality in scientific research},\ }\href
  {https://doi.org/10.1007/BF02016604} {\bibfield  {journal} {\bibinfo
  {journal} {Scientometrics}\ }\textbf {\bibinfo {volume} {9}},\ \bibinfo
  {pages} {13} (\bibinfo {year} {1986})}\BibitemShut {NoStop}%
\bibitem [{\citenamefont {Simonton}(1988)}]{simonton1988scientific}%
  \BibitemOpen
  \bibfield  {author} {\bibinfo {author} {\bibfnamefont {D.~K.}\ \bibnamefont
  {Simonton}},\ }\href {https://doi.org/10.1002/acp.2350050210} {\emph
  {\bibinfo {title} {Scientific genius: {A} psychology of science}}}\ (\bibinfo
   {publisher} {Cambridge University Press},\ \bibinfo {year}
  {1988})\BibitemShut {NoStop}%
\bibitem [{\citenamefont {Feist}(1997)}]{feist1997quantity}%
  \BibitemOpen
  \bibfield  {author} {\bibinfo {author} {\bibfnamefont {G.~J.}\ \bibnamefont
  {Feist}},\ }\bibfield  {title} {\bibinfo {title} {Quantity, quality, and
  depth of research as influences on scientific eminence: {Is} quantity most
  important?},\ }\href {https://doi.org/10.1207/s15326934crj1004_4} {\bibfield
  {journal} {\bibinfo  {journal} {Creativity Research Journal}\ }\textbf
  {\bibinfo {volume} {10}},\ \bibinfo {pages} {325} (\bibinfo {year}
  {1997})}\BibitemShut {NoStop}%
\bibitem [{\citenamefont {Haslam}\ and\ \citenamefont
  {Laham}(2010)}]{haslam2010quality}%
  \BibitemOpen
  \bibfield  {author} {\bibinfo {author} {\bibfnamefont {N.}~\bibnamefont
  {Haslam}}\ and\ \bibinfo {author} {\bibfnamefont {S.~M.}\ \bibnamefont
  {Laham}},\ }\bibfield  {title} {\bibinfo {title} {Quality, quantity, and
  impact in academic publication},\ }\href {https://doi.org/10.1002/ejsp.727}
  {\bibfield  {journal} {\bibinfo  {journal} {European Journal of Social
  Psychology}\ }\textbf {\bibinfo {volume} {40}},\ \bibinfo {pages} {216}
  (\bibinfo {year} {2010})}\BibitemShut {NoStop}%
\bibitem [{\citenamefont {Nijstad}\ \emph {et~al.}(2010)\citenamefont
  {Nijstad}, \citenamefont {Dreu}, \citenamefont {Rietzschel},\ and\
  \citenamefont {Baas}}]{nijstad2010dual}%
  \BibitemOpen
  \bibfield  {author} {\bibinfo {author} {\bibfnamefont {B.~A.}\ \bibnamefont
  {Nijstad}}, \bibinfo {author} {\bibfnamefont {C.~K.~D.}\ \bibnamefont
  {Dreu}}, \bibinfo {author} {\bibfnamefont {E.~F.}\ \bibnamefont
  {Rietzschel}},\ and\ \bibinfo {author} {\bibfnamefont {M.}~\bibnamefont
  {Baas}},\ }\bibfield  {title} {\bibinfo {title} {The dual pathway to
  creativity model: Creative ideation as a function of flexibility and
  persistence},\ }\href {https://doi.org/10.1080/10463281003765323} {\bibfield
  {journal} {\bibinfo  {journal} {European Review of Social Psychology}\
  }\textbf {\bibinfo {volume} {21}},\ \bibinfo {pages} {34} (\bibinfo {year}
  {2010})}\BibitemShut {NoStop}%
\bibitem [{\citenamefont {Bosquet}\ and\ \citenamefont
  {Combes}(2013)}]{bosquet2013academics}%
  \BibitemOpen
  \bibfield  {author} {\bibinfo {author} {\bibfnamefont {C.}~\bibnamefont
  {Bosquet}}\ and\ \bibinfo {author} {\bibfnamefont {P.-P.}\ \bibnamefont
  {Combes}},\ }\bibfield  {title} {\bibinfo {title} {Are academics who publish
  more also more cited? {Individual} determinants of publication and citation
  records},\ }\href {https://doi.org/10.1007/s11192-013-0996-6} {\bibfield
  {journal} {\bibinfo  {journal} {Scientometrics}\ }\textbf {\bibinfo {volume}
  {97}},\ \bibinfo {pages} {831} (\bibinfo {year} {2013})}\BibitemShut
  {NoStop}%
\bibitem [{\citenamefont {Abramo}\ \emph {et~al.}(2014)\citenamefont {Abramo},
  \citenamefont {Cicero},\ and\ \citenamefont
  {D’Angelo}}]{abramo2014authors}%
  \BibitemOpen
  \bibfield  {author} {\bibinfo {author} {\bibfnamefont {G.}~\bibnamefont
  {Abramo}}, \bibinfo {author} {\bibfnamefont {T.}~\bibnamefont {Cicero}},\
  and\ \bibinfo {author} {\bibfnamefont {C.~A.}\ \bibnamefont {D’Angelo}},\
  }\bibfield  {title} {\bibinfo {title} {Are the authors of highly cited
  articles also the most productive ones?},\ }\href
  {https://doi.org/10.1016/j.joi.2013.10.011} {\bibfield  {journal} {\bibinfo
  {journal} {Journal of Informetrics}\ }\textbf {\bibinfo {volume} {8}},\
  \bibinfo {pages} {89} (\bibinfo {year} {2014})}\BibitemShut {NoStop}%
\bibitem [{\citenamefont {Sandstr{\"o}m}\ and\ \citenamefont {van~den
  Besselaar}(2016)}]{sandstrom2016quantity}%
  \BibitemOpen
  \bibfield  {author} {\bibinfo {author} {\bibfnamefont {U.}~\bibnamefont
  {Sandstr{\"o}m}}\ and\ \bibinfo {author} {\bibfnamefont {P.}~\bibnamefont
  {van~den Besselaar}},\ }\bibfield  {title} {\bibinfo {title} {Quantity and/or
  quality? {The} importance of publishing many papers},\ }\href
  {https://doi.org/10.1371/journal.pone.0166149} {\bibfield  {journal}
  {\bibinfo  {journal} {PLoS ONE}\ }\textbf {\bibinfo {volume} {11}},\ \bibinfo
  {pages} {e0166149} (\bibinfo {year} {2016})}\BibitemShut {NoStop}%
\bibitem [{\citenamefont {Larivi{\`e}re}\ and\ \citenamefont
  {Costas}(2016)}]{lariviere2016many}%
  \BibitemOpen
  \bibfield  {author} {\bibinfo {author} {\bibfnamefont {V.}~\bibnamefont
  {Larivi{\`e}re}}\ and\ \bibinfo {author} {\bibfnamefont {R.}~\bibnamefont
  {Costas}},\ }\bibfield  {title} {\bibinfo {title} {How many is too many? {On}
  the relationship between research productivity and impact},\ }\href
  {https://doi.org/10.1371/journal.pone.0162709} {\bibfield  {journal}
  {\bibinfo  {journal} {PLoS ONE}\ }\textbf {\bibinfo {volume} {11}},\ \bibinfo
  {pages} {e0162709} (\bibinfo {year} {2016})}\BibitemShut {NoStop}%
\bibitem [{\citenamefont {Garousi}\ and\ \citenamefont
  {Fernandes}(2017)}]{garousi2017quantity}%
  \BibitemOpen
  \bibfield  {author} {\bibinfo {author} {\bibfnamefont {V.}~\bibnamefont
  {Garousi}}\ and\ \bibinfo {author} {\bibfnamefont {J.~M.}\ \bibnamefont
  {Fernandes}},\ }\bibfield  {title} {\bibinfo {title} {Quantity versus impact
  of software engineering papers: {A} quantitative study},\ }\href
  {https://doi.org/10.1007/s11192-017-2419-6} {\bibfield  {journal} {\bibinfo
  {journal} {Scientometrics}\ }\textbf {\bibinfo {volume} {112}},\ \bibinfo
  {pages} {963} (\bibinfo {year} {2017})}\BibitemShut {NoStop}%
\bibitem [{\citenamefont {Michalska-Smith}\ and\ \citenamefont
  {Allesina}(2017)}]{michalska2017and}%
  \BibitemOpen
  \bibfield  {author} {\bibinfo {author} {\bibfnamefont {M.~J.}\ \bibnamefont
  {Michalska-Smith}}\ and\ \bibinfo {author} {\bibfnamefont {S.}~\bibnamefont
  {Allesina}},\ }\bibfield  {title} {\bibinfo {title} {And, not or: {Quality},
  quantity in scientific publishing},\ }\href
  {https://doi.org/10.1371/journal.pone.0178074} {\bibfield  {journal}
  {\bibinfo  {journal} {PLoS ONE}\ }\textbf {\bibinfo {volume} {12}},\ \bibinfo
  {pages} {e0178074} (\bibinfo {year} {2017})}\BibitemShut {NoStop}%
\bibitem [{\citenamefont {Kolesnikov}\ \emph {et~al.}(2018)\citenamefont
  {Kolesnikov}, \citenamefont {Fukumoto},\ and\ \citenamefont
  {Bozeman}}]{kolesnikov2018researchers}%
  \BibitemOpen
  \bibfield  {author} {\bibinfo {author} {\bibfnamefont {S.}~\bibnamefont
  {Kolesnikov}}, \bibinfo {author} {\bibfnamefont {E.}~\bibnamefont
  {Fukumoto}},\ and\ \bibinfo {author} {\bibfnamefont {B.}~\bibnamefont
  {Bozeman}},\ }\bibfield  {title} {\bibinfo {title} {Researchers’
  risk-smoothing publication strategies: Is productivity the enemy of
  impact?},\ }\href {https://doi.org/10.1007/s11192-018-2793-8} {\bibfield
  {journal} {\bibinfo  {journal} {Scientometrics}\ }\textbf {\bibinfo {volume}
  {116}},\ \bibinfo {pages} {1995} (\bibinfo {year} {2018})}\BibitemShut
  {NoStop}%
\bibitem [{\citenamefont {Bornmann}\ and\ \citenamefont
  {Tekles}(2019)}]{bornmann2019productivity}%
  \BibitemOpen
  \bibfield  {author} {\bibinfo {author} {\bibfnamefont {L.}~\bibnamefont
  {Bornmann}}\ and\ \bibinfo {author} {\bibfnamefont {A.}~\bibnamefont
  {Tekles}},\ }\bibfield  {title} {\bibinfo {title} {Productivity does not
  equal usefulness},\ }\href {https://doi.org/10.1007/s11192-018-2982-5}
  {\bibfield  {journal} {\bibinfo  {journal} {Scientometrics}\ }\textbf
  {\bibinfo {volume} {118}},\ \bibinfo {pages} {705} (\bibinfo {year}
  {2019})}\BibitemShut {NoStop}%
\bibitem [{\citenamefont {Forthmann}\ \emph {et~al.}(2020)\citenamefont
  {Forthmann}, \citenamefont {Leveling}, \citenamefont {Dong},\ and\
  \citenamefont {Dumas}}]{forthmann2020investigating}%
  \BibitemOpen
  \bibfield  {author} {\bibinfo {author} {\bibfnamefont {B.}~\bibnamefont
  {Forthmann}}, \bibinfo {author} {\bibfnamefont {M.}~\bibnamefont {Leveling}},
  \bibinfo {author} {\bibfnamefont {Y.}~\bibnamefont {Dong}},\ and\ \bibinfo
  {author} {\bibfnamefont {D.}~\bibnamefont {Dumas}},\ }\bibfield  {title}
  {\bibinfo {title} {Investigating the quantity--quality relationship in
  scientific creativity: {An} empirical examination of expected residual
  variance and the tilted funnel hypothesis},\ }\href
  {https://doi.org/10.1007/s11192-020-03571-w} {\bibfield  {journal} {\bibinfo
  {journal} {Scientometrics}\ }\textbf {\bibinfo {volume} {124}},\ \bibinfo
  {pages} {2497} (\bibinfo {year} {2020})}\BibitemShut {NoStop}%
\bibitem [{\citenamefont {Larivi{\`e}re}\ and\ \citenamefont
  {Sugimoto}(2019)}]{lariviere2019jifhistory}%
  \BibitemOpen
  \bibfield  {author} {\bibinfo {author} {\bibfnamefont {V.}~\bibnamefont
  {Larivi{\`e}re}}\ and\ \bibinfo {author} {\bibfnamefont {C.~R.}\ \bibnamefont
  {Sugimoto}},\ }\bibinfo {title} {The journal impact factor: {A} brief
  history, critique, and discussion of adverse effects}\ (\bibinfo  {publisher}
  {Springer},\ \bibinfo {year} {2019})\ pp.\ \bibinfo {pages}
  {3--24}\BibitemShut {NoStop}%
\bibitem [{\citenamefont {McKiernan}\ \emph {et~al.}(2019)\citenamefont
  {McKiernan}, \citenamefont {Schimanski}, \citenamefont {Nieves},
  \citenamefont {Matthias}, \citenamefont {Niles},\ and\ \citenamefont
  {Alperin}}]{mckiernan2019meta}%
  \BibitemOpen
  \bibfield  {author} {\bibinfo {author} {\bibfnamefont {E.~C.}\ \bibnamefont
  {McKiernan}}, \bibinfo {author} {\bibfnamefont {L.~A.}\ \bibnamefont
  {Schimanski}}, \bibinfo {author} {\bibfnamefont {C.~M.}\ \bibnamefont
  {Nieves}}, \bibinfo {author} {\bibfnamefont {L.}~\bibnamefont {Matthias}},
  \bibinfo {author} {\bibfnamefont {M.~T.}\ \bibnamefont {Niles}},\ and\
  \bibinfo {author} {\bibfnamefont {J.~P.}\ \bibnamefont {Alperin}},\
  }\bibfield  {title} {\bibinfo {title} {Meta-research: {Use} of the journal
  impact factor in academic review, promotion, and tenure evaluations},\ }\href
  {https://doi.org/10.7554/eLife.47338} {\bibfield  {journal} {\bibinfo
  {journal} {eLife}\ }\textbf {\bibinfo {volume} {8}},\ \bibinfo {pages}
  {e47338} (\bibinfo {year} {2019})}\BibitemShut {NoStop}%
\bibitem [{\citenamefont {Bornmann}\ and\ \citenamefont
  {Leydesdorff}(2017)}]{bornmann2017skewness}%
  \BibitemOpen
  \bibfield  {author} {\bibinfo {author} {\bibfnamefont {L.}~\bibnamefont
  {Bornmann}}\ and\ \bibinfo {author} {\bibfnamefont {L.}~\bibnamefont
  {Leydesdorff}},\ }\bibfield  {title} {\bibinfo {title} {Skewness of citation
  impact data and covariates of citation distributions: {A} large-scale
  empirical analysis based on {Web of Science} data},\ }\href
  {https://doi.org/10.1016/j.joi.2016.12.001} {\bibfield  {journal} {\bibinfo
  {journal} {Journal of Informetrics}\ }\textbf {\bibinfo {volume} {11}},\
  \bibinfo {pages} {164} (\bibinfo {year} {2017})}\BibitemShut {NoStop}%
\bibitem [{\citenamefont {Traag}(2021)}]{traag2021inferring}%
  \BibitemOpen
  \bibfield  {author} {\bibinfo {author} {\bibfnamefont {V.~A.}\ \bibnamefont
  {Traag}},\ }\bibfield  {title} {\bibinfo {title} {Inferring the causal effect
  of journals on citations},\ }\href {https://doi.org/10.1162/qss_a_00128}
  {\bibfield  {journal} {\bibinfo  {journal} {Quantitative Science Studies}\
  }\textbf {\bibinfo {volume} {2}},\ \bibinfo {pages} {496} (\bibinfo {year}
  {2021})}\BibitemShut {NoStop}%
\bibitem [{\citenamefont {Kim}\ \emph {et~al.}(2020)\citenamefont {Kim},
  \citenamefont {Portenoy}, \citenamefont {West},\ and\ \citenamefont
  {Stovel}}]{kim2019scientific}%
  \BibitemOpen
  \bibfield  {author} {\bibinfo {author} {\bibfnamefont {L.}~\bibnamefont
  {Kim}}, \bibinfo {author} {\bibfnamefont {J.~H.}\ \bibnamefont {Portenoy}},
  \bibinfo {author} {\bibfnamefont {J.~D.}\ \bibnamefont {West}},\ and\
  \bibinfo {author} {\bibfnamefont {K.~W.}\ \bibnamefont {Stovel}},\ }\bibfield
   {title} {\bibinfo {title} {Scientific journals still matter in the era of
  academic search engines and preprint archives},\ }\href
  {https://doi.org/10.1002/asi.24326} {\bibfield  {journal} {\bibinfo
  {journal} {Journal of the Association for Information Science and
  Technology}\ }\textbf {\bibinfo {volume} {71}},\ \bibinfo {pages} {1218}
  (\bibinfo {year} {2020})}\BibitemShut {NoStop}%
\bibitem [{\citenamefont {Correa}\ \emph {et~al.}(2021)\citenamefont {Correa},
  \citenamefont {Laverde-Rojas}, \citenamefont {Tejada},\ and\ \citenamefont
  {Marmolejo-Ramos}}]{correa2020scihub}%
  \BibitemOpen
  \bibfield  {author} {\bibinfo {author} {\bibfnamefont {J.~C.}\ \bibnamefont
  {Correa}}, \bibinfo {author} {\bibfnamefont {H.}~\bibnamefont
  {Laverde-Rojas}}, \bibinfo {author} {\bibfnamefont {J.}~\bibnamefont
  {Tejada}},\ and\ \bibinfo {author} {\bibfnamefont {F.}~\bibnamefont
  {Marmolejo-Ramos}},\ }\bibfield  {title} {\bibinfo {title} {The {Sci-Hub}
  effect on papers’ citations},\ }\bibfield  {journal} {\bibinfo  {journal}
  {Scientometrics}\ }\href {https://doi.org/10.1007/s11192-020-03806-w}
  {10.1007/s11192-020-03806-w} (\bibinfo {year} {2021})\BibitemShut {NoStop}%
\bibitem [{\citenamefont {Waltman}\ and\ \citenamefont
  {Traag}(2020)}]{waltman2020use}%
  \BibitemOpen
  \bibfield  {author} {\bibinfo {author} {\bibfnamefont {L.}~\bibnamefont
  {Waltman}}\ and\ \bibinfo {author} {\bibfnamefont {V.~A.}\ \bibnamefont
  {Traag}},\ }\bibfield  {title} {\bibinfo {title} {Use of the journal impact
  factor for assessing individual articles need not be statistically wrong},\
  }\href {https://doi.org/10.12688/f1000research.23418.1} {\bibfield  {journal}
  {\bibinfo  {journal} {F1000Research}\ }\textbf {\bibinfo {volume} {9}},\
  \bibinfo {pages} {366} (\bibinfo {year} {2020})}\BibitemShut {NoStop}%
\bibitem [{SI()}]{SI}%
  \BibitemOpen
  \href@noop {} {\bibinfo {title} {{Supplemental Material for Figs. S1--S21,
  and Tables S1 and S2.}}}\BibitemShut {Stop}%
\bibitem [{\citenamefont {de~Solla~Price}(1963)}]{solla1963little}%
  \BibitemOpen
  \bibfield  {author} {\bibinfo {author} {\bibfnamefont {D.~J.}\ \bibnamefont
  {de~Solla~Price}},\ }\href@noop {} {\emph {\bibinfo {title} {Little
  {Science}, {Big} {Science}}}}\ (\bibinfo  {publisher} {Columbia University
  Press},\ \bibinfo {year} {1963})\BibitemShut {NoStop}%
\bibitem [{\citenamefont {Sinatra}\ \emph {et~al.}(2016)\citenamefont
  {Sinatra}, \citenamefont {Wang}, \citenamefont {Deville}, \citenamefont
  {Song},\ and\ \citenamefont {Barab{\'a}si}}]{sinatra2016quantifying}%
  \BibitemOpen
  \bibfield  {author} {\bibinfo {author} {\bibfnamefont {R.}~\bibnamefont
  {Sinatra}}, \bibinfo {author} {\bibfnamefont {D.}~\bibnamefont {Wang}},
  \bibinfo {author} {\bibfnamefont {P.}~\bibnamefont {Deville}}, \bibinfo
  {author} {\bibfnamefont {C.}~\bibnamefont {Song}},\ and\ \bibinfo {author}
  {\bibfnamefont {A.-L.}\ \bibnamefont {Barab{\'a}si}},\ }\bibfield  {title}
  {\bibinfo {title} {Quantifying the evolution of individual scientific
  impact},\ }\href {https://doi.org/10.1126/science.aaf5239} {\bibfield
  {journal} {\bibinfo  {journal} {Science}\ }\textbf {\bibinfo {volume}
  {354}},\ \bibinfo {pages} {aaf5239} (\bibinfo {year} {2016})}\BibitemShut
  {NoStop}%
\bibitem [{\citenamefont {Antonoyiannakis}(2018)}]{antonoyiannakis2018impact}%
  \BibitemOpen
  \bibfield  {author} {\bibinfo {author} {\bibfnamefont {M.}~\bibnamefont
  {Antonoyiannakis}},\ }\bibfield  {title} {\bibinfo {title} {Impact factors
  and the {Central Limit Theorem}: {Why} citation averages are scale
  dependent},\ }\href {https://doi.org/10.1016/j.joi.2018.08.011} {\bibfield
  {journal} {\bibinfo  {journal} {Journal of Infometrics}\ }\textbf {\bibinfo
  {volume} {12}},\ \bibinfo {pages} {1072} (\bibinfo {year}
  {2018})}\BibitemShut {NoStop}%
\bibitem [{\citenamefont {Antonoyiannakis}(2020)}]{antonoyiannakis2020impact}%
  \BibitemOpen
  \bibfield  {author} {\bibinfo {author} {\bibfnamefont {M.}~\bibnamefont
  {Antonoyiannakis}},\ }\bibfield  {title} {\bibinfo {title} {Impact factor
  volatility due to a single paper: {A} comprehensive analysis},\ }\href
  {https://doi.org/10.1162/qss_a_00037} {\bibfield  {journal} {\bibinfo
  {journal} {Quantitative Science Studies}\ }\textbf {\bibinfo {volume} {1}},\
  \bibinfo {pages} {639} (\bibinfo {year} {2020})}\BibitemShut {NoStop}%
\bibitem [{\citenamefont {Hicks}(2012)}]{hicks2012performance}%
  \BibitemOpen
  \bibfield  {author} {\bibinfo {author} {\bibfnamefont {D.}~\bibnamefont
  {Hicks}},\ }\bibfield  {title} {\bibinfo {title} {Performance-based
  university research funding systems},\ }\href
  {https://doi.org/10.1016/j.respol.2011.09.007} {\bibfield  {journal}
  {\bibinfo  {journal} {Research Policy}\ }\textbf {\bibinfo {volume} {41}},\
  \bibinfo {pages} {251} (\bibinfo {year} {2012})}\BibitemShut {NoStop}%
\bibitem [{\citenamefont {Price}(2018)}]{price2018seventy}%
  \BibitemOpen
  \bibfield  {author} {\bibinfo {author} {\bibfnamefont {M.}~\bibnamefont
  {Price}},\ }\href {https://doi.org/doi:10.1126/science.aav4004} {\bibinfo
  {title} {{Some scientists publish more than 70 papers a year. Here’s how --
  and why -- they do it}}} (\bibinfo {year} {2018}),\ \bibinfo {note} {accessed
  September 2020}\BibitemShut {NoStop}%
\bibitem [{Note1()}]{Note1}%
  \BibitemOpen
  \bibinfo {note} {\protect \url {http://lattes.cnpq.br/}}\BibitemShut
  {NoStop}%
\bibitem [{\citenamefont {Garfield}(1972)}]{garfield1972citation}%
  \BibitemOpen
  \bibfield  {author} {\bibinfo {author} {\bibfnamefont {E.}~\bibnamefont
  {Garfield}},\ }\bibfield  {title} {\bibinfo {title} {Citation analysis as a
  tool in journal evaluation: {Journals} can be ranked by frequency and impact
  of citations for science policy studies},\ }\href
  {https://doi.org/10.1126/science.178.4060.471} {\bibfield  {journal}
  {\bibinfo  {journal} {Science}\ }\textbf {\bibinfo {volume} {178}},\ \bibinfo
  {pages} {471} (\bibinfo {year} {1972})}\BibitemShut {NoStop}%
\bibitem [{\citenamefont {Garfield}(2006)}]{garfield2006if}%
  \BibitemOpen
  \bibfield  {author} {\bibinfo {author} {\bibfnamefont {E.}~\bibnamefont
  {Garfield}},\ }\bibfield  {title} {\bibinfo {title} {{The history and meaning
  of the journal impact factor}},\ }\href
  {https://doi.org/10.1001/jama.295.1.90} {\bibfield  {journal} {\bibinfo
  {journal} {Journal of the American Medical Association}\ }\textbf {\bibinfo
  {volume} {295}},\ \bibinfo {pages} {90} (\bibinfo {year} {2006})}\BibitemShut
  {NoStop}%
\bibitem [{\citenamefont {Guerrero-Bote}\ and\ \citenamefont
  {Moya-Anegón}(2012)}]{guerrerobote2012sjr2}%
  \BibitemOpen
  \bibfield  {author} {\bibinfo {author} {\bibfnamefont {V.~P.}\ \bibnamefont
  {Guerrero-Bote}}\ and\ \bibinfo {author} {\bibfnamefont {F.}~\bibnamefont
  {Moya-Anegón}},\ }\bibfield  {title} {\bibinfo {title} {A further step
  forward in measuring journals{'} scientific prestige: The {SJR2} indicator},\
  }\href {https://doi.org/10.1016/j.joi.2012.07.001} {\bibfield  {journal}
  {\bibinfo  {journal} {Journal of Informetrics}\ }\textbf {\bibinfo {volume}
  {6}},\ \bibinfo {pages} {674} (\bibinfo {year} {2012})}\BibitemShut {NoStop}%
\bibitem [{\citenamefont {Bordons}\ \emph {et~al.}(2002)\citenamefont
  {Bordons}, \citenamefont {Fern{\'a}ndez},\ and\ \citenamefont
  {G{\'o}mez}}]{bordons2002advantages}%
  \BibitemOpen
  \bibfield  {author} {\bibinfo {author} {\bibfnamefont {M.}~\bibnamefont
  {Bordons}}, \bibinfo {author} {\bibfnamefont {M.}~\bibnamefont
  {Fern{\'a}ndez}},\ and\ \bibinfo {author} {\bibfnamefont {I.}~\bibnamefont
  {G{\'o}mez}},\ }\bibfield  {title} {\bibinfo {title} {Advantages and
  limitations in the use of impact factor measures for the assessment of
  research performance},\ }\href {https://doi.org/10.1023/A:1014800407876}
  {\bibfield  {journal} {\bibinfo  {journal} {Scientometrics}\ }\textbf
  {\bibinfo {volume} {53}},\ \bibinfo {pages} {195} (\bibinfo {year}
  {2002})}\BibitemShut {NoStop}%
\bibitem [{\citenamefont {Foster}\ \emph {et~al.}(2015)\citenamefont {Foster},
  \citenamefont {Rzhetsky},\ and\ \citenamefont {Evans}}]{foster2015tradition}%
  \BibitemOpen
  \bibfield  {author} {\bibinfo {author} {\bibfnamefont {J.~G.}\ \bibnamefont
  {Foster}}, \bibinfo {author} {\bibfnamefont {A.}~\bibnamefont {Rzhetsky}},\
  and\ \bibinfo {author} {\bibfnamefont {J.~A.}\ \bibnamefont {Evans}},\
  }\bibfield  {title} {\bibinfo {title} {Tradition and innovation in
  scientists{\textquoteright} research strategies},\ }\href
  {https://doi.org/10.1177/0003122415601618} {\bibfield  {journal} {\bibinfo
  {journal} {American Sociological Review}\ }\textbf {\bibinfo {volume} {80}},\
  \bibinfo {pages} {875} (\bibinfo {year} {2015})}\BibitemShut {NoStop}%
\bibitem [{\citenamefont {Huber}(2004)}]{huber2004robust}%
  \BibitemOpen
  \bibfield  {author} {\bibinfo {author} {\bibfnamefont {P.~J.}\ \bibnamefont
  {Huber}},\ }\href@noop {} {\emph {\bibinfo {title} {Robust Statistics}}}\
  (\bibinfo  {publisher} {Wiley},\ \bibinfo {year} {2004})\BibitemShut
  {NoStop}%
\bibitem [{sta(2020)}]{statsmodels}%
  \BibitemOpen
  \href@noop {} {\bibinfo {title} {{statsmodels}}} (\bibinfo {year} {2020}),\
  \bibinfo {note} {available at \url{https://www.statsmodels.org}. Accessed
  August 2020.}\BibitemShut {Stop}%
\bibitem [{\citenamefont {Gelman}(2006)}]{Gelman06priordistributions}%
  \BibitemOpen
  \bibfield  {author} {\bibinfo {author} {\bibfnamefont {A.}~\bibnamefont
  {Gelman}},\ }\bibfield  {title} {\bibinfo {title} {Prior distributions for
  variance parameters in hierarchical models},\ }\href
  {https://doi.org/10.1214/06-BA117A} {\bibfield  {journal} {\bibinfo
  {journal} {Bayesian Analysis}\ }\textbf {\bibinfo {volume} {1}},\ \bibinfo
  {pages} {515} (\bibinfo {year} {2006})}\BibitemShut {NoStop}%
\bibitem [{\citenamefont {Salvatier}\ \emph {et~al.}(2016)\citenamefont
  {Salvatier}, \citenamefont {Wiecki},\ and\ \citenamefont
  {Fonnesbeck}}]{pymc3}%
  \BibitemOpen
  \bibfield  {author} {\bibinfo {author} {\bibfnamefont {J.}~\bibnamefont
  {Salvatier}}, \bibinfo {author} {\bibfnamefont {T.~V.}\ \bibnamefont
  {Wiecki}},\ and\ \bibinfo {author} {\bibfnamefont {C.}~\bibnamefont
  {Fonnesbeck}},\ }\bibfield  {title} {\bibinfo {title} {Probabilistic
  programming in {Python} using {PyMC3}},\ }\href
  {https://doi.org/10.7717/peerj-cs.55} {\bibfield  {journal} {\bibinfo
  {journal} {PeerJ Computer Science}\ }\textbf {\bibinfo {volume} {2}},\
  \bibinfo {pages} {e55} (\bibinfo {year} {2016})}\BibitemShut {NoStop}%
\end{thebibliography}%

\clearpage
\includepdf[pages=1-22,pagecommand={\thispagestyle{empty}, \clearpage}]{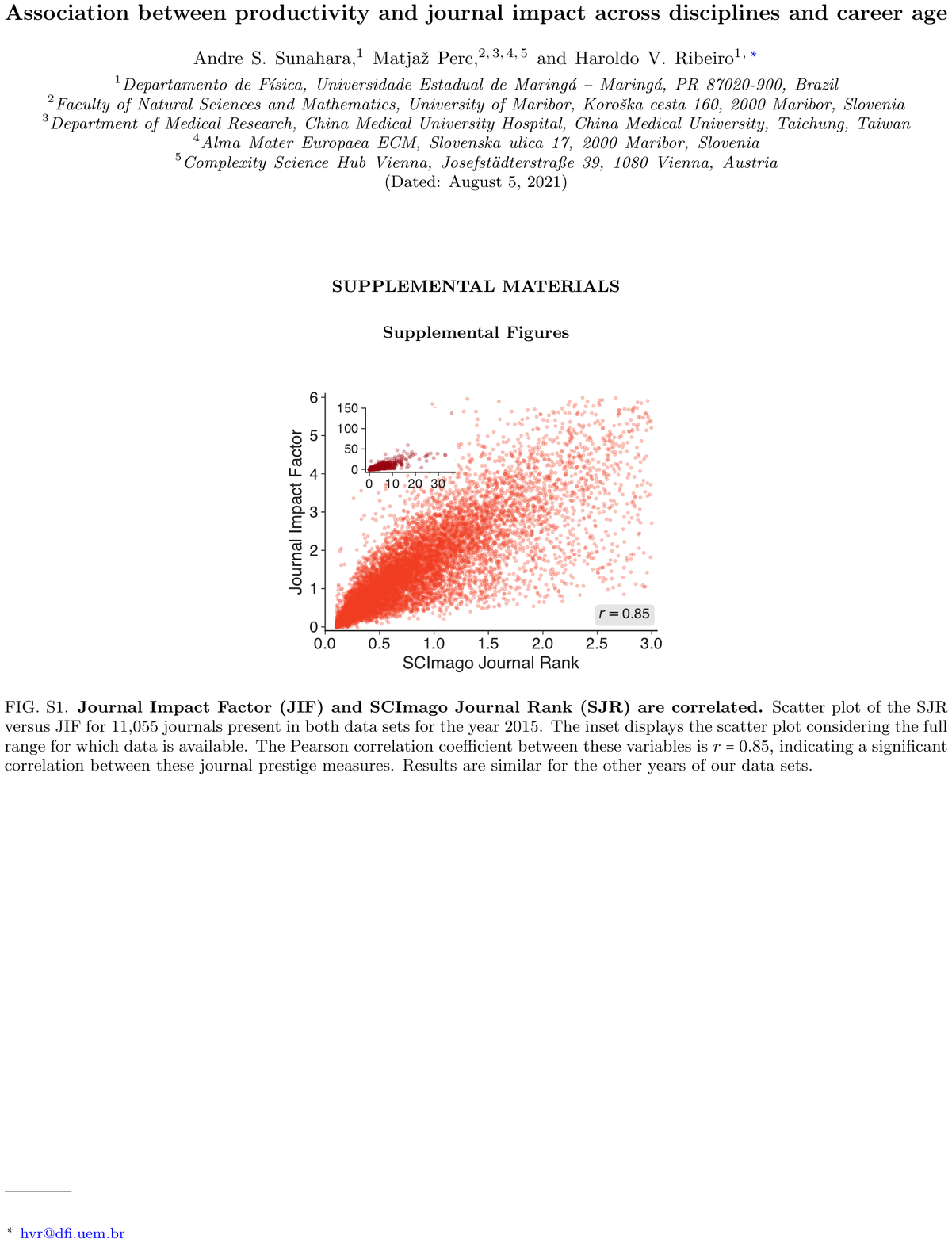}
\includepdf[pages=23,pagecommand={\thispagestyle{empty}}]{prr_supplementary.pdf}

\end{document}